\documentclass{aa}
\usepackage{graphics,graphicx}
\usepackage[varg]{txfonts}
\usepackage{rotating}
\usepackage{lscape}
\usepackage{rotating}
\usepackage{natbib}
\usepackage{longtable}
\newcommand{\asec}{$^{\prime\prime}$}
\newcommand{\pas}{.\hskip-2pt$^{\prime\prime}$}

\def\OMC{OMC-2 FIR4}

\def\CIII{HC$_3$N}
\def\CV{HC$_5$N}

\def\kms{\mbox{km~s$^{-1}$}}

\def\cmq{cm$^{-2}$}

\def\solm{\mbox{M$_\odot$}}
\def\soll{\mbox{L$_\odot$}}

\begin{document}
\title{SOLIS II. Carbon-chain growth in the Solar-type protocluster
  OMC2-FIR4
\thanks{Based on observations carried out under project number L15AA with the IRAM NOEMA Interferometer.
IRAM is supported by INSU/CNRS (France), MPG (Germany) and IGN (Spain).}}
\author{F. Fontani\inst{1}
         \and C. Ceccarelli\inst{2} 
         \and C. Favre\inst{2}
         \and P. Caselli\inst{3}
         \and R. Neri\inst{4}
         \and I.R. Sims\inst{5}
         \and C. Kahane\inst{2}
         \and F. Alves\inst{3}
         \and N. Balucani\inst{6}
         \and E. Bianchi\inst{1,7}
         \and E. Caux\inst{8,9}
         \and A. Jaber Al-Edhari\inst{2, 10}
         \and A. Lopez-Sepulcre\inst{4}
         \and J. E. Pineda\inst{3}
         \and R. Bachiller\inst{11}
         \and L. Bizzocchi\inst{3}
         \and S. Bottinelli\inst{8,9}
         \and A. Chacon-Tanarro\inst{3}
         \and R. Choudhury\inst{3}
         \and C. Codella\inst{1}
         \and A. Coutens\inst{12}
         \and F. Dulieu\inst{13}
         \and S. Feng\inst{3}
         \and A. Rimola\inst{14}
         \and P. Hily-Blant\inst{2}
         \and J. Holdship\inst{12}
         \and I. Jimenez-Serra\inst{12,15}
         \and J. Laas\inst{3}
         \and B. Lefloch\inst{2}
         \and Y. Oya\inst{16}
         \and L. Podio\inst{1}
         \and A. Pon\inst{17}
         \and A. Punanova\inst{3}
         \and D. Quenard\inst{15}
         \and N. Sakai\inst{18}
         \and S. Spezzano\inst{3}
         \and V. Taquet\inst{19}
         \and L. Testi\inst{1,20}
         \and P. Theul\'e\inst{21}
         \and P. Ugliengo\inst{22}
         \and C. Vastel\inst{7,8}
         \and A.I. Vasyunin\inst{3,23}
         \and S. Viti\inst{12}
         \and S. Yamamoto\inst{16}
         \and L. Wiesenfeld\inst{2}
         }     
\offprints{F. Fontani, \email{fontani@arcetri.astro.it}}
\institute{INAF-Osservatorio Astrofisico di Arcetri, Largo E. Fermi 5, I-50125, Florence, Italy
          \and 
          Univ. Grenoble Alpes, CNRS, IPAG, F-38000 Grenoble, France 
          \and
          Max-Planck-Institut f\"ur extraterrestrische Physik (MPE), D-85748 Garching, Germany 
          \and
          Institut de Radioastronomie Millim\'etrique, 300 rue de la Piscine, 38406, Saint-Martin d'H\`eres, France
          \and
          Institut de Physique de Rennes, UMR CNRS 6251, Universit\'e de Rennes 1, 263 Avenue du G\'en\'eral Leclerc, F-35042, Rennes Cedex, France
          \and
          Dipartimento di Chimica, Biologia e Biotecnologie, Universit\`a di Perugia, Via Elce di Sotto 8, I-06123 Perugia, Italy
          \and
          Dipartimento di Fisica e Astronomia, Universit\`a degli Studi di Firenze, I-50125 Firenze, Italy 
          \and
          Universit\'e de Toulouse, UPS-OMP, IRAP, Toulouse, France
          \and 
          CNRS, IRAP, 9 Av. Colonel Roche, BP 44346, F-31028 Toulouse Cedex 4, France
          \and
          University of AL-Muthanna, College of Science, Physics Department, AL-Muthanna, Iraq
          \and 
          Observatorio Astron\'omico Nacional (OAN, IGN), Calle Alfonso XII, 3, 28014 Madrid, Spain
          \and
          Department of Physics and Astronomy, University College London, Gower St., London, WC1E 6BT, UK
          \and
          LERMA, Universit\'e de Cergy-Pontoise, Observatoire de Paris, PSL Research University, CNRS, Sorbonne Universit\'es, UPMC Univ. Paris 06, 95000 Cergy Pontoise, France
          \and
          Departament de Qu\'imica, Universitat Aut\`onoma de Barcelona, E-08193 Bellaterra, Spain
          \and
          Astronomy Unit, School of Physics \& Astronomy, Queen Mary University of London, Mile End Road, London E1 4NS, UK
          \and
          Department of Physics, The University of Tokyo, Bunkyo-ku, Tokyo 113-0033, Japan
          \and
          Department of Physics and Astronomy, The University of Western Ontario, 1151 Richmond Street, London, N6A 3K7, Canada
          \and 
          The Institute of Physical and Chemical Research (RIKEN), 2-1, Hirosawa, Wako-shi, Saitama 351-0198, Japan
          \and
          Leiden Observatory, Leiden University, P.O. Box 9513, 2300-RA Leiden, The Netherlands
          \and
          European Southern Observatory, Karl-Schwarzschild-Str. 2, 85748 Garching bei M\"{u}nchen, Germany
          \and
          Aix-Marseille Universit\'e, PIIM UMR-CNRS 7345, 13397 Marseille, France
          \and
          Dipartimento di Chimica and NIS Centre, Universit\`a degli Studi di Torino, Via P. Giuria 7, I-10125 Torino, Italy
	  \and
	  Ural Federal University, Ekaterinburg, Russia
          }
\date{Received date; accepted date}

\titlerunning{Carbon-chain growth in OMC2}
\authorrunning{Fontani et al.}


\abstract{The interstellar delivery of carbon atoms locked into
molecules might be one of the key ingredients for the emergence of
life. Cyanopolyynes are carbon chains delimited at their two
extremities by an atom of hydrogen and a cyano group, so that they
might be excellent reservoirs of carbon. The simplest member, HC$_3$N,
is ubiquitous in the galactic interstellar medium and found also in
external galaxies.  Thus, understanding the growth of cyanopolyynes
in regions forming stars similar to our Sun, and what affects it, is
particularly relevant. In the framework of the IRAM/NOEMA Large
Program SOLIS (Seeds Of Life In Space), we have obtained a map of
two cyanopolyynes, \CIII\ and \CV\, in the protocluster \OMC . 
Because our Sun is thought to be born in a rich cluster, \OMC\ is
one of the closest and best known representatives of the environment
in which the Sun may have been born. We find a \CIII /\CV\ abundance
ratio across the source in the range $\sim 1 - 30$, with the
smallest values ($\leq 10$) in FIR5 and in the Eastern region of
FIR4. The ratios $\leq 10$ can be reproduced by chemical models only
if: (1) the cosmic-ray ionisation rate $\zeta$ is
$\sim 4\times 10^{-14}$ s$^{-1}$; (2) the gaseous elemental ratio
C/O is close to unity; (3) O and C are largely
depleted. The large $\zeta$ is comparable to that measured in FIR4
by previous works and was interpreted as due to a flux of energetic
($\geq 10$ MeV) particles from embedded sources. We suggest that
these sources could lie East of FIR4 and FIR5. 
A temperature gradient across FIR4, with $T$ decreasing 
by about 10 K, could also explain the observed change in 
the \CIII /\CV\ line ratio, without the need of a cosmic ray ionisation rate gradient. 
However, even in this case, a high constant cosmic-ray ionisation
rate (of the order of $10^{-14}$ s$^{-1}$) is necessary to reproduce the observations.} 

\keywords{Stars: formation -- ISM: clouds -- ISM: molecules -- Radio lines: ISM}

\maketitle
%
\section{Introduction}
\label{intro}

The origin of life, as we know it, requires the simultaneous presence
of at least two "ingredients": liquid water and carbon atoms. In the
past years, a lot of work has been devoted to the search for water
reservoirs in all the evolutionary steps that lead to the formation of
Sun-like stars, from pre-stellar cores (Caselli et
al.~\citeyear{caselli12}), to accreting protostars (Ceccarelli et
al. 1999, J{o}rgensen et al.~\citeyear{jorgensen}), to protoplanetary
disks (Hogerheijde et al.~\citeyear{hogerheijde}, 
Podio et al.~\citeyear{podio13}, Cleeves et al.~\citeyear{cleeves}). 
However, relatively little is known about the
presence and growth of carbon chains, which might be an important
reservoir of carbon atoms usable to build large biotic molecules
(e.g.~Loison et al.~\citeyear{loison}, Balucani et al.~\citeyear{balucani2000},
Balucani~\citeyear{balucani2009}). In
this respect, cyanopolyynes, i.e. carbon chain molecules with an 
atom of hydrogen and a group of CN at their extremities (generic
formula: HC$_{\rm 2n+1}$N), are among the best species to
understand the formation of carbon chains. Indeed, they are ubiquitous
in the interstellar medium, detected in the Milky Way as well as in
external galaxies (e.g. Broten et al.~\citeyear{broten}, Bell et al.~\citeyear{bell92},
Mauersberger et al.~\citeyear{mauersberger}).  Even more interesting in the
astrobiological context, cyanoplyynes were detected in protoplanetary
disks (Chapillon et al.~\citeyear{chapillon}, \"{O}berg et al.~\citeyear{oberg}), 
on Titan's atmosphere (e.g.~Vuitton et al.~\citeyear{vuitton})
and comets (e.g.~Mumma \& Charnley~\citeyear{mec2011}),
the continuous rain of which may have enriched the primitive Earth of
carbon usable for synthesising biological molecules.  All this makes
cyanopolyynes excellent potential reservoirs of molecular carbon for
the formation of longer chains of (pre-)biological importance.

The largest cyanopolyyne detected so far in the interstellar medium is
HC$_{9}$N (Broten et al.~\citeyear{broten}). However, only relatively small
cyanopolyynes, up to HC$_7$N, have been detected in Solar-type
protostars and in a relatively small sample: L1521E (Hirota et 
al.~\citeyear{hirota}), L1527 (Sakai et al.~\citeyear{sakai08}), L1512 
and L1251A (Cordiner et al.~\citeyear{cordiner11}), Cha-MMS1 
(Cordiner et al.~\citeyear{cordiner12}), 
and IRAS16293-2422 (Jaber Al-Edhari et al.~\citeyear{jaber}). Hence, what makes
cyanopolyynes thrive in Solar-type protostellar environments is still
an open question. In this paper, we report observations of \CV\ and \CIII\ towards the
source \OMC\, obtained with the interferometer NOEMA within the Large
Program SOLIS (Seeds Of Life In Space: Ceccarelli et al. in prep.). 

\OMC\ is part of the Orion Molecular Cloud 2, north of the famous KL
object, at a distance of $\sim$420 pc (Menten et
al.~\citeyear{menten}; Hirota et al.~\citeyear{hirota07}). It lies on
a bridge of material, probably a filament where new protostars are
forming (Chini et al.~\citeyear{chini}). FIR4 is in between two other
young protostars: FIR3 (also known as SOF 2N or HOPS 370: Adams et
al.~\citeyear{adams12}), about 30$"$ north-west, and FIR5 (SOF 4 or
HOPS 369: Adams et al.~\citeyear{adams12}), about 20$"$ south-east
(Mezger et al.~\citeyear{mezger}). \OMC\ is itself a young
proto-cluster which harbours several embedded low- and
intermediate-mass protostars (Shimajiri et al.~\citeyear{shimajiri},
L\'opez-Sepulcre et al.~\citeyear{lopez13}).  High-velocity symmetric
wings in high-J CO and water lines detected by Herschel suggest the
presence of a yet unveiled compact outflow from one of the embedded
sources of the FIR4 cluster. The total mass of FIR4 is
around 30 \solm\ (Mezger et al.~\citeyear{mezger}; Crimier et
  al.~\citeyear{crimier}) and its luminosity is less than 1000 \soll\ (Crimier et
  al.~\citeyear{crimier}, Furlan et al.~\citeyear{furlan}), respectively. Adams et
  al. (2012) found a mid-IR source, SOF 3 or HOPS 108, whose bolometric 
  luminosity is estimated to be 38~\soll\ by Furlan et al.~(\citeyear{furlan}), 
  associated with the extended FIR4 millimeter emission, but not coincident with its
  emission peak. No specific
  observations focused on FIR3 and FIR5, the status of which,
  therefore, remains poorly known. Finally, Herschel
observations have revealed the presence of an inner source of
energetic ($\geq$10 MeV) particles which ionise the surrounding
envelope up to a distance of 3500--5000 AU (Ceccarelli et
al.~\citeyear{ceccarelli14}). Intriguingly, the derived irradiation
dose is of the same order of magnitude as that experienced by the
young Sun and whose trace is recorded in meteoritic material, by the
so-called Short-Lived Radionuclides (SLRs: e.g. Gounelle et
al.~\citeyear{gounelle}). This and the mounting evidence that the Sun
was born in a crowded cluster of stars (even including massive stars)
rather than in an isolated clump (Adams~\citeyear{adams}) make \OMC\
the closest analogue of what must have been the environment of
our Sun at the very beginning of its formation. In this context, the study of the
cyanopolyynes towards \OMC\ provides constraints on their growth
in a similar natal environment.

\section{Observations and results}
\label{obs}

Observations with the IRAM NOEMA Interferometer of \CIII\ (9--8) and
\CV\ (31-30), at rest frequencies 81.881468~GHz (Thorwirth et 
al.~\citeyear{thorwirth}) and 82.539039~GHz (Bizzocchi et 
al.~\citeyear{bizzocchi04}), respectively, towards \OMC\ have been carried 
out in 5 days between the 5th and the 19th of August, 2015. We used the 
array in D configuration, with baselines from 15 to 95 m, providing an angular
resolution of $\sim 9.5\times 6.1$\asec\ at the frequency of both
lines. The phase center was R.A.(J2000)= 05$^h$35$^m$26\pas 97,
Dec.(J2000)= --05$^{\circ}$09$^{\prime}$56\pas 8, and the local
standard of rest velocity was set to $11.4$ \kms . The primary beam is
$\sim 61$\asec .  The system temperature was in between 100 and 200~K
in all tracks, and the amount of precipitable water vapour was
generally around 10~mm.  The calibration of the bandpass was performed
on 3C454.3, while 0524+034 was used for calibration of the gains in
phase and amplitude.  The absolute flux scale was set by observing
MWC349 ($\sim 1.0$~Jy at 82.1~GHz) when observable (three out of five
days), and LKHA101 ($\sim 0.21$~Jy at 82.1~GHz) during the other days.

The \CIII\ line was observed in the Widex band correlator, providing a
resolution in velocity of $\sim 7.15$ \kms , while the \CV\
line was observed also in the Narrow band correlator with a resolution in
velocity of $\sim$0.57 \kms . The following analysis of the \CV\ line will be
conducted using mainly the higher resolution spectrum. 
We will show in the following that the the difference in the spectral 
resolutions has a negligible effect on our analysis, which is focused on 
comparing the integrated intensities of the two lines.
The continuum was imaged by averaging the line-free channels of the Widex 
and Narrow correlator units. Calibration and imaging were performed using 
the CLIC and MAPPING softwares of the GILDAS\footnote{The GILDAS 
software is developed at the IRAM and the Observatoire de Grenoble, 
and is available at http://www.iram.fr/IRAMFR/GILDAS} package using 
standard procedures. The continuum image was self-calibrated, and the 
solutions were applied to the lines. Other lines detected in the Widex 
broad band correlator will be presented in accompanying papers.

The line strength and energy of the upper level of the two transitions are given
in Table~\ref{tab_res}, and are taken from the 
Cologne Database for Molecular Spectroscopy (CDMS; M\"{u}ller et 
al.~\citeyear{muller01}, M\"{u}ller et al.~\citeyear{muller05}).

\begin{figure}
\begin{center}
\includegraphics[angle=0,width=9cm]{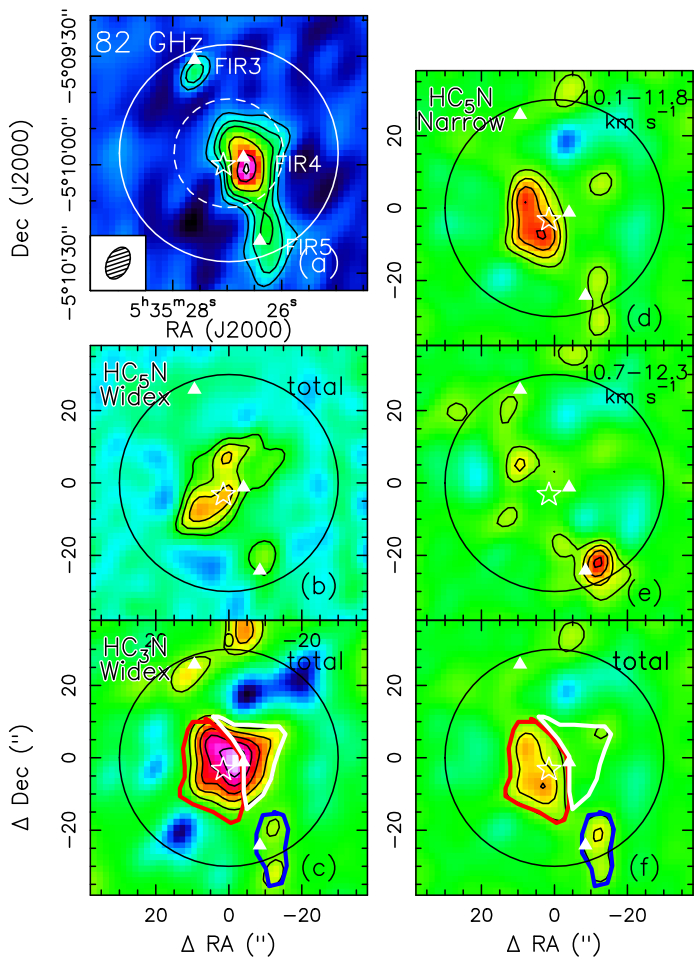}
\caption[]
{\label{fig_maps}{{\it Panel (a):} 82~GHz continuum emission obtained with
IRAM NOEMA towards \OMC . The first contour level is at 1.8$\times 10^{-3}$
Jy beam$^{-1}$, corresponding to the 3$\sigma$ rms level of the map,
and the others are 2, 3, 5, 8, and 10 times the 3$\sigma$ rms level.
The white solid circle shows the NOEMA field of view, while the dashed
one is the IRAM-30m HPBW ($\sim $30\asec ). The triangles indicate the
position of FIR3, FIR4 and FIR5, while the open star corresponds to
the mid-IR protostar detected at 8, 24, and 70 $\mu$m by Furlan 
et al.~(\citeyear{furlan}).
{\it Panel (b):} integrated intensity map of \CV\ (31--30) obtained from the
Widex correlator integrated in the velocity channels with signal.
The contour levels are 3, 5, and 7 times the $1\sigma$ rms of the map, 
equal to $\sim 8.9\times 10^{-4}$ Jy beam$^{-1}$.
The solid circle (in this panel and in all panels from {\it (b)}
to {\it (f)}) represents the NOEMA field of view (equal to the
white circle in panel {\it (a)}.
{\it Panel (c):} integrated intensity map of \CIII\ (9--8). Both the first level and step
correspond to the $3\sigma$ rms value of the integrated map (1$\sigma
\sim 6.3\times 10^{-3}$ Jy beam$^{-1}$).
{\it Panels (d, e, f)}: map of \CV\ (31--30) obtained averaging the emission 
in the three different velocity ranges indicated in the top right corner
extracted from the Narrow correlator unit.
In all panels, both the first contour level and step correspond to the 
$3\sigma$ rms value of the total integrated map, showed in panel 
{\it (f)} (1$\sigma \sim 3.6\times 10^{-3}$ Jy beam$^{-1}$). 
Finally, in {\it panels (c)} and {\it (f)}, 
the red and white contours correspond to the emitting region of \CIII\
towards FIR4 with high and low \CV\ emission, respectively. 
From these contours, we have extracted the spectra discussed in
Sect.~\ref{discu}.
The blue contour indicates FIR5 as detected in the total integrated map 
of \CV\ (panel {\it (f)}).
}
}
\end{center}
\end{figure}

From the presented dataset, we have detected both the \CIII\ (9--8)
and \CV\ (31--30) transitions towards \OMC\ with excellent signal-to-noise 
ratio (larger than 10 for both lines). Fig.~\ref{fig_maps} shows the 
morphology of the continuum emission at $\sim 82$~GHz (panel {\it (a)}), 
and that of the velocity-averaged emission of both lines in panels {\it (b) - (f)}. 
In Fig.~\ref{fig_maps}, panels {\it (b)} and {\it (f)} show the emission map
of \CV\ (9--8) obtained by integrating the line over the channels with signal 
(in the Widex and Narrow spectra, respectively). Both plots indicate that the 
\CV\ emission arises mainly from two condensations associated with FIR4 
and FIR5.

In panel {\it (c)}, we show the integrated emission of \CIII\ (9--8), which also 
arises mainly from FIR4 and FIR5, but significant emission is detected also
towards FIR3. Overall, the continuum emission is more 
consistent with that of \CIII\ (compare panels {\it (a)} and {\it (c)}) 
than with that of \CV\ (compare panels {\it (a)} and {\it (b) or (f)}), 
although in both lines the emission is clearly shifted (and more
extended) to the East of the continuum peak. 

Concerning the comparison between the two carbon-chain species, 
the \CIII\ emission is well centred towards FIR4, while the \CV\ emission 
is clearly displaced East of the map centre (compare panel {\it (c)} with panel 
{\it (b)} or {\it (f)}).
Towards FIR5, both lines show faint and compact emission. 
Interestingly, the \CV\ emission seems clearly associated with the mid-IR 
embedded protostar detected at 8, 24, and 70 $\mu$m by Furlan et 
al.~(\citeyear{furlan}), indicated by the open star in Fig.~\ref{fig_maps}. 
This suggests an enhanced formation of \CV\ close to this protostar. Moreover,
there is possibly a link between the different distribution of the two cyanopolyynes
and the free-free emission detected with the JVLA (Lopez-Sepulcre, pers. comm.)
East of FIR4. 
We will come back more extensively to this point in Sect.~\ref{discu}.
However, a comparison with the JVLA data goes beyond the scope of
this work and will be discussed in a forthcoming paper.

In both lines, we resolve out a significant fraction of the extended flux.
Fig.~\ref{flux_comp} shows the comparison between the NOEMA and IRAM-30m
spectra extracted from the IRAM-30m beam: we find that in FIR4 we recover only $20\%$ 
of the total flux in \CV , and $\sim 15\%$ in \CIII . Therefore, even though our maps 
are certainly resolving out a lot of extended emission, the fraction of this emission
is comparable (within the calibration errors), and it is related to the same angular 
scale (because the observations of the two lines are simultaneous).
Therefore, the comparison between the two maps should be only marginally affected 
by the differential filtering.
Details on how we have estimated the missing flux are given in 
Sect.~\ref{appa_missingflux} of the Appendix.

As stated before, our analysis on \CV\ is mainly based on the
higher-spectral resolution data extracted from the Narrow correlator.
However, we have evaluated 
whether the different spectral resolution had an effect in the comparison 
of the two lines: for this purpose, we have also extracted the spectrum of the
\CV\ line from the Widex correlator. As an example, in Fig.~\ref{flux_comp}, 
we show the comparison between the Narrow and Widex spectra of \CV\ 
extracted from the region FIR4-\CV\ (red contour in Fig.~\ref{fig_maps}). 
We have found that the difference in the integrated intensity between the
Widex and Narrow spectrum is only $\sim 7\%$, smaller than the calibration 
uncertainty. 
Thus, because our analysis will be based on the total integrated intensity
of the two lines, our conclusions will not be not affected by the 
different spectral resolutions.
Moreover, by comparing the maps of the total integrated emission of \CV\
obatined from the Widex (panel {\it (b)}) and the Narrow (panel {\it (f)})
correlators, it is apparent that the two correlators do not produce maps with 
major overall differences in the morphology of the \CV\ emission.

Due to the low spectral resolution of the \CIII\ observations
($\sim 7.15$ \kms ), it is impossible to derive any kinematical
information from this line.  On the other hand, \CV\ (31--30) was
observed with a spectral resolution of $\sim 0.5$ \kms , which allows
us to spectrally resolve the line. We find that the bulk of the
emission towards FIR4 is at a velocity of $\sim 10.9$ \kms\ ({\it panel
 d}), while that towards FIR5 is at $\sim 11.5$ \kms\ ({\it panel
 e}), indicating a small but significant difference in the 
velocities of the two condensations (the systemic one being
$\sim 11.4$ \kms , e.g. Lopez-Sepulcre et al.~\citeyear{lopez13}). 
However, as shown in panels {\it (b)} and
{\it (f)} of Fig.~\ref{fig_maps}, smoothing the \CV\ line to
the same spectral resolution of the \CIII\ does not significantly 
change the morphology of the line emission.
The same effect is thus expected for the \CIII , which allows us
to conclude that the fact that \CIII\ is more extended to the West 
than \CV\ is not due to its lower spectral resolution (and thus higher 
sensitivity).
One can reasonably expect the same effect on the \CIII\ line, because
the \CV\ line shape of the spectra extracted from Widex is similar to that 
of \CIII . As an example, in Fig.~\ref{flux_comp} we show this comparison
for the spectra of the two lines extracted in the region FIR4-\CV .

Because the velocity-averaged emission of \CIII\ and \CV\ show morphological
differences, we have extracted the spectra of both molecules from three sub-regions, 
indicated in panels {\it (b)} and {\it (f)} of Fig.~\ref{fig_maps} by a red, a blue, and
a white contour. The red and blue contours represent the regions of FIR4 and FIR5,
respectively, in which the velocity-averaged emission of both \CIII\ and \CV\ 
are above 3$\sigma$ rms.
The white contour instead indicates the regions of FIR4 where the velocity-averaged
emission of \CIII\ is above 3$\sigma$ rms, while that of \CV\ is under this value.
The integrated flux densities of the lines extracted from these three regions, as
well as those extracted from the total FIR4 region, are reported in Table~\ref{tab_res}.

In summary, the analysis reported above shows that:\\
(i) \CIII\ emission is mostly associated with FIR4 with a weak
peak associated with FIR3 and FIR5;\\
(ii) \CV\ emission is also associated mainly with FIR4, but a strong
peak is also associated with FIR5;\\
(iii) \CIII\ and \CV\ emission do not spatially coincide in FIR4:
specifically, while \CIII\ emission overlaps quite well with the continuum
emission, \CV\ emits only in the eastern-half of it.

\section{Modeling}
\label{model}

\subsection{\CIII/\CV\ abundance ratio}\label{sec:deriv-ciiicv-abund}
As mentioned in the Introduction, the main goal of this work is to
understand the cyanopolyyne formation/growth. We, therefore, start
analysing the \CIII/\CV\ intensity ratio in the three regions reported
in Table~\ref{tab_res}, FIR4-\CV\ (red contour in
Fig.~\ref{fig_maps}), FIR4-low\CV\ (white contour), and FIR5 (blue
contour), to estimate the \CIII/\CV\ abundance ratio in each of
them. To this end, we computed the \CIII/\CV\ line intensity ratio
assuming LTE and optically thin lines, as a function of the gas
temperature and assuming various \CIII/\CV\ abundance ratios. Note
that the estimated gas temperature in the FIR4 extended envelope is
35--45 K (Ceccarelli et al.~\citeyear{ceccarelli14}), whereas no
estimates of the gas temperature exist towards FIR5.

\begin{figure}
\begin{center}
  \includegraphics[angle=0,width=9cm]{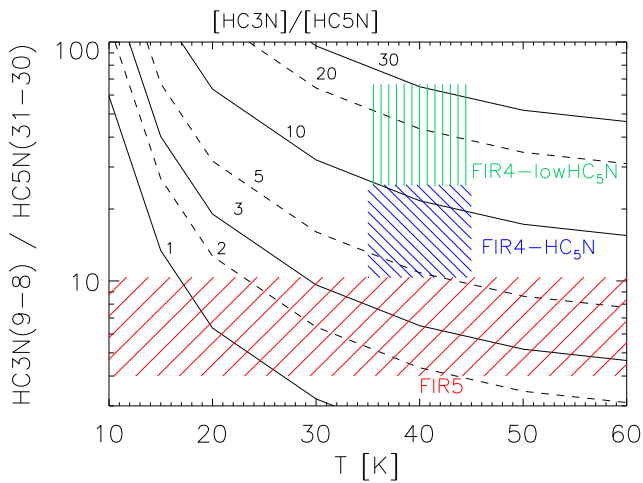}
  \caption[]{\label{fig_ratios_1}{Theoretical \CIII\ (9-8)/\CV\ (31-30) line
      intensity ratio assuming LTE and optically thin conditions as a
      function of the gas temperature for different values
      of the \CIII /\CV\ abundance ratio as marked (black curves). The range
      of values observed within a 2 $\sigma$ error bar towards the three regions listed in
      Tab. \ref{tab_res} are showed as dashed areas: FIR4-\CV\ (blue),
      FIR4-low\CV\ (green) and FIR5 (red). Note that the temperature of
      FIR4 was estimated to be $\sim 40$~K
      (Ceccarelli et al.~\citeyear{ceccarelli14}) while no estimates
      for FIR5 exist.}}
\end{center}
\end{figure}

The results of the optically thin LTE modeling and their comparison
with the observed line intensity ratios are reported in
Fig.~\ref{fig_ratios_1} and Table~\ref{tab_res}.  They imply that: (a)
in FIR4-\CV\, the \CIII/\CV\ abundance ratio is between 4 and 12; (b)
in FIR4-low\CV\, the \CIII/\CV\ abundance ratio is between 10 and 30; (c)
in FIR5, the \CIII/\CV\ abundance ratio is less than 6, regardless of
the the gas temperature (within 10 and 60 K).

Note that single-dish observations towards OMC-2 FIR4 show that the
two lines are optically thin or only moderately optically thick
(Jaber Al-Edhari et al. in preparation). In Appendix~\ref{appc}, 
we show the IRAM-30m spectra of the \CIII\ (9--8) line and its three 
$^{13}$C isotopologues, from which we have deduced the opacity 
of the main isotopologue line.
Concerning the possible non-LTE excitation, since non-LTE effects
would be much more severe for the \CV\ line than the \CIII\ line (the
\CV\ line has the higher upper level energy, and therefore it is more difficult 
to populate it according to LTE), the \CIII/\CV\ abundance ratio found by
the LTE analysis could be overestimated.

Another possible explanation for the observed difference in the
  \CIII /\CV\ line intensity ratio is a slight temperature gradient
  across FIR4.  Assuming that the average temperature of FIR4 is about 40~K
  (Ceccarelli et al. 2014), if one allows the gradient to be about
  10~K, with the Eastern region FIR4-\CV\ being the warmer
  (Fig.~\ref{fig_ratios_1}), this leads to two possibilities. A) In
  FIR4-\CV\ the temperature is at most as high as $\sim 50$~K: the
  \CIII/\CV\ would be in the $4-14$ range in FIR4-\CV , while it
  remains $10-30$ in FIR4-low\CV . In this case, the ratio might be
  constant, $10-14$, across the whole region.  B) In FIR4-low\CV\ the
  temperature is as low as 30~K: the \CIII/\CV\ remains in the $4-12$
  range in FIR4-\CV , while it would be $8-30$ in FIR4-low\CV . In
  this case the ratio may be constant, around $8-12$, across the
  region.  Therefore, while a temperature gradient could explain the
  observed line ratio gradient, it would imply a \CIII/\CV\ abundance
  ratio lower than 14.
Please note that, since the amount of missing flux is comparable in the two 
lines (see Sect.~\ref{obs}), our column density ratios are expected to be 
affected by this by less than 30\%. 

Finally, we estimated the abundance of HC$_3$N 
in the three regions illustrated in Fig.~\ref{fig_maps}
  (FIR4-\CV , FIR4-low\CV , and FIR5) from the line integrated
  intensity and continuum emission. The total column density of \CIII\
  was calculated assuming LTE for the line population, optically thin
  conditions, and gas temperatures between 10 and 50 K. The H$_2$
  column densities, $N({\rm H_2})$, were derived from the dust mass
  computed from the 3~mm continuum emission (panel {\it (a)} in
  Fig.~\ref{fig_maps}) from the equation:
\begin{equation} 
M_{\rm dust}=\frac{S_{\nu}d^2}{\kappa_{\nu}B_{\nu}(T)}
\end{equation}
where: $S_{\nu}$ is the total integrated flux density; $d$ is the
source distance; $\kappa_{\nu}$ is the dust mass opacity coefficient,
extrapolated at 3~mm from the value of 1 cm$^2$ gr$^{-1}$
at 250~GHz (Ossenkopf \& Henning~\citeyear{oeh}), assuming a
dust opacity index $\beta = 2$, i.e. a spectral index $2+\beta=4$; 
and $B_{\nu}(T)$ is the Planck function at dust temperature $T$. The
equation is valid for optically thin emission. Then, we have computed
the H$_2$ mass by multiplying $M_{\rm dust}$ for a mass gas-to-dust
ratio of 100, from which we have computed the average H$_2$ volume
density assuming a spherical source, and finally computed
$N({\rm H_2})$ by multiplying the volume density for the average
diameter of the sources.  We used dust temperatures in the range
10--50~K, as for the gas.  We have obtained the following
$N({\rm H_2})$ values: $\sim 1-5 \times 10^{23}$ \cmq\ in FIR4-\CV ;
$\sim 2-9\times 10^{23}$ \cmq\ in FIR4-low\CV ; and
$\sim 1.5-7\times 10^{23}$ \cmq\ in FIR5.  The resulting HC$_3$N
abundance with respect to H atoms is in the range 0.5 -- 5 $\times 10^{-11}$.

\begin{table*}
\begin{center}
\caption[] {Line spectroscopic parameters, and integrated flux density of the 
\CIII\ (9--8) and \CV\ (31--30) lines extracted from the regions indicated in 
Fig.~\ref{fig_maps}. Uncertainties are in brackets and include
the calibration error on the absolute flux scale of the order of $\sim 10\%$. 
Last row reports the calculated HC$_3$N/HC$_5$N abundance ratio, 
as computed in Sec.~\ref{model}.}
\label{tab_res}
\small
\begin{tabular}{lcccccc}
\hline \hline
Line & $E_{\rm u}$ & $S \mu^2$ &  \multicolumn{4}{c}{Integrated flux density}  \\
       \cline{4-7} \\
     & & & FIR4-total & FIR4-\CV\ (red) & FIR4-low\CV\ (white) & FIR5 (blue) \\
     & K & D$^2$ & Jy \kms\  & Jy \kms\  &  Jy \kms\  &  Jy \kms\  \\
\hline
HC$_3$N(9--8) & 19.6 & 124.8 & 6.0(0.6)     & 4.0(0.4) & 2.0(0.2) & 0.46(0.05) \\
HC$_5$N(31--30) & 63.4 & 581 & 0.29(0.03) & 0.25(0.03) & 0.050(0.007) & 0.072(0.009) \\
\hline
HC$_3$N/HC$_5$N & & &   & 4-12   & 10-30 & $\leq 6$ \\
\hline
\end{tabular}
\end{center}
\end{table*}
\normalsize

\subsection{Chemistry}
\label{chemistry}

Large cyanopolyynes are commonly associated with the
early chemical evolution of molecular clouds, when carbon atoms are
not yet completely locked up into CO (e.g. Loison et
al.~\citeyear{loison} and references therein) and, hence, no protostar
is yet present. An important chemical process relevant for the cyanopolyynes in
protostars is the differential freeze-out of light versus heavy
molecules, with the former sticking onto the dust grains
faster/earlier than the latter, which would introduce a so-called
``freeze-out peak'' of cyanopolyyne abundance in the gas (e.g., Brown
\& Charnley~\citeyear{bec}). Furthermore, the sublimation of methane
from ices, when the dust temperature exceeds about 30 K, introduces
into the gas-phase carbon atoms for reactions leading to cyanopolyynes
(Sakai et al.~\citeyear{sakai08}, Hassel et al.~\citeyear{hassel}).
Grain surface chemistry may also play an important role in the evolution 
of the abundances of cyanopolyynes (see Graninger et al.~\citeyear{graninger}).

Based on the analysis made in Sect.~\ref{model}, the cyanopolyynes 
growth appears different in the three regions of Table \ref{tab_res}, with the \CIII /\CV\
abundance ratio being smaller ($\leq 6$) in FIR5, larger (10--30) in
FIR4-low\CV\ and in between (4-12) in FIR4-\CV. In order to understand
the origin of this difference, we run a time-dependent astrochemical
model with different parameters. We used a modified
version of Nahoon with an upgraded version of the chemical network
KIDA\footnote{The original network and code are publicly available at
  {\it http://kida.obs.u-bordeaux1.fr} (Wakelam et
  al.~\citeyear{wakelam2014}).}. The modifications of Nahoon are to
improve its usage flexibility, while the modifications of KIDA take
into account the upgrade of the carbon chains chemistry by Loison et
al.~(\citeyear{loison}), new reactions by Balucani et
al.~(\citeyear{balucani}) and Barone et al.~(\citeyear{barone}), and
new values for the reactions CN + C$_4$H$_2$ $\rightarrow$ HC$_5$N +
H, C$_2$H + HC$_3$N $\rightarrow$ HC$_5$N + H, and C$_3$N + C$_2$H$_2$
$\rightarrow$ HC$_5$N + H, following laboratory experiments and
computations by Cheikh~(\citeyear{cheikh}), Fournier (\citeyear{fournier14}) and 
Fournier et al. in prep. Details on these reactions are reported in Appendix~\ref{appb}.

The code was ran assuming that H$_2$ is initially molecular, while the
other elements are either ionised or atomic. Note that all abundances
are given with respect to H nuclei. Gaseous oxygen $A_O$ and carbon
$A_C$ elemental abundances are varied from 0.5 to $2 \times 10^{-4}$ 
and 0.3 to $1.5 \times 10^{-4}$, respectively, to simulate the freezing-out of
these two elements into the grain mantles.  We assumed that the
nitrogen abundance $A_N$ scales by the same factor as carbon, since
CO and N$_2$, the major carbon and nitrogen reservoirs in molecular
gas, have the same binding energies (e.g. Bisschop et al.~\citeyear{bisschop}; 
Noble et al.~\citeyear{noble}).  We therefore computed $A_N$ according to the formula
$6.2 \times 10^{-5} \times A_C /1.7\times 10^{-4}$. The other elements
are depleted by a factor 100 with respect to the Solar photosphere
values following Jenkins et al.~(\citeyear{jenkins}), as in several
similar modelling works (e.g. Wakelam et al.~\citeyear{wakelam2010};
Kong et al.~\citeyear{kong}).
We assumed the gas temperature and density are equal to those measured in
the envelope of FIR4, namely T$\sim40$ K and n$_{H2}\sim 1.2 \times 10^6$ cm$^{-3}$ 
(Ceccarelli et al.~\citeyear{ceccarelli14}). 
For FIR5 we used the same values. For the temperature, this assumption is
justified by the study of Friesen et al.~(\citeyear{friesen17}) who have found, 
from ammonia measurements, that the temperature changes by a few degrees 
in the whole region encompassing FIR3, FIR4 and FIR5.
%
Finally, we ran three grids of models with the cosmic-ray ionisation
rate $\zeta$ equal to $1\times10^{-17}$, $3\times10^{-16}$ and
$4\times10^{-14}$ s$^{-1}$, respectively. The first represents the
``canonical'' value for galactic molecular clouds (e.g. Padovani et
al. 2009), the second one is the value measured in the diffuse cloud
in the vicinity of FIR4 (Lopez-Sepulcre et al.~\citeyear{lopez13b}), and the last
one is the value measured in the envelope of FIR4 (Ceccarelli et
al.~\citeyear{ceccarelli14}).

\begin{figure*}
\begin{center}
\includegraphics[angle=0,width=9cm]{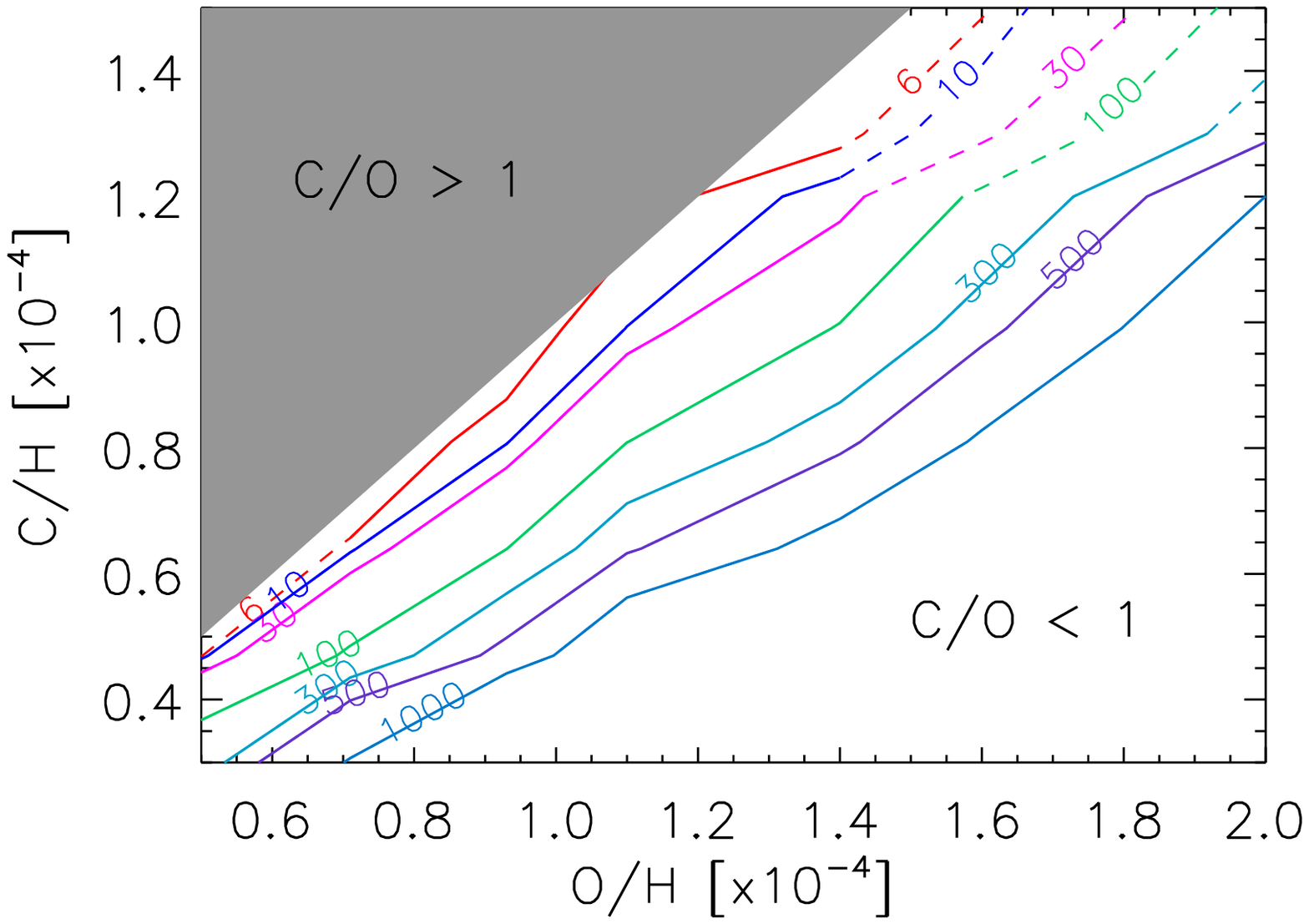}
\includegraphics[angle=0,width=9cm]{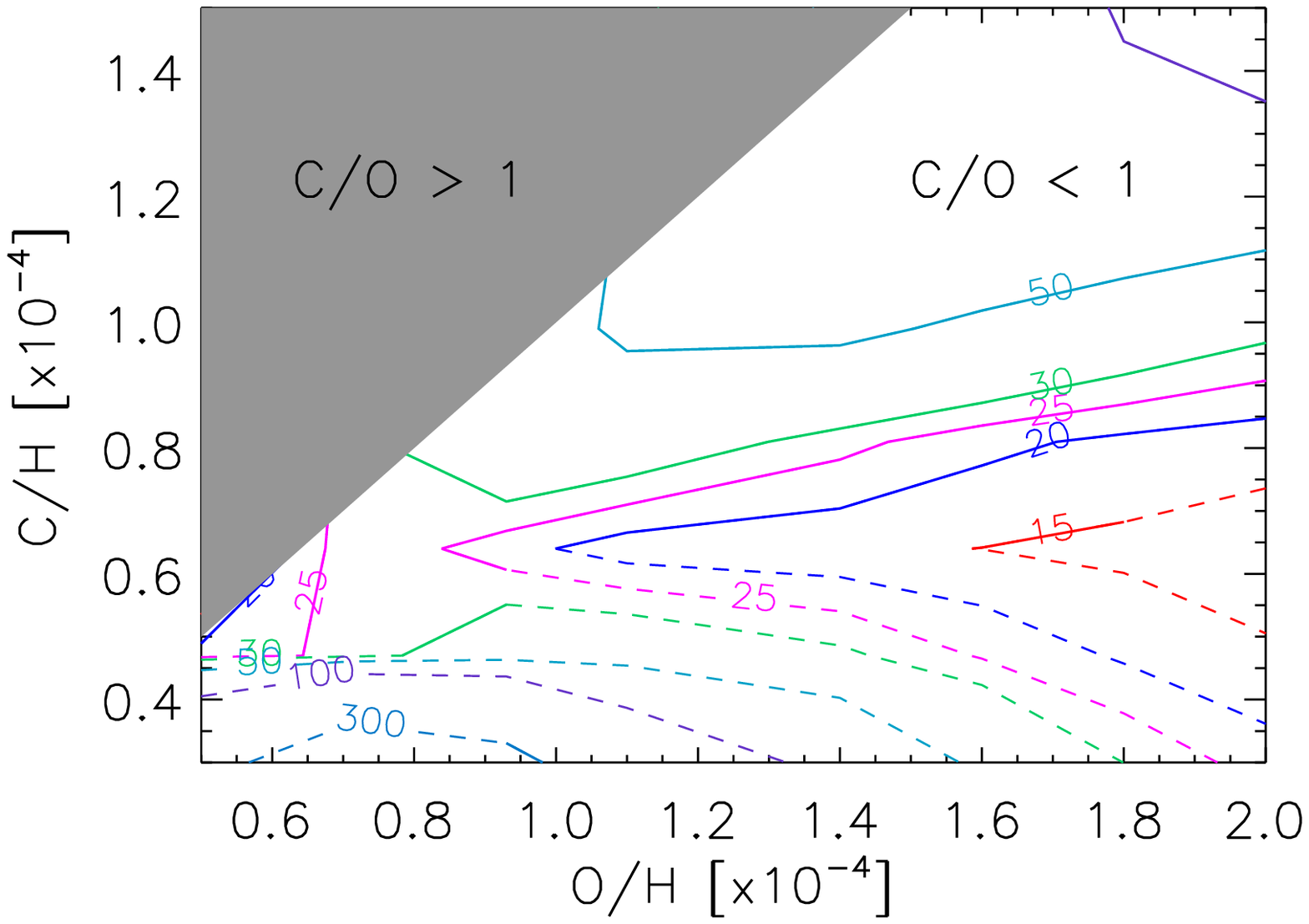}
\caption[] {\label{fig_ratios} Contour plots of the \CIII/\CV\
  abundance ratio as a function of the O/H and C/H gaseous elemental
  abundance. The grey zone marks the region where the C/O elemental ratio
  is $\geq1$, which is unrealistic (see text). The solid lines mark
  the loci where the HC$_3$N abundance is equal to the measured
  one, namely between 0.5 and 5 $\times 10^{-11}$. The models were
  obtained for a gas temperature of 40 K and 
  a H$_2$ density of $1.2 \times 10^6$ cm$^{-3}$, valid for the
  envelope of FIR4 (Ceccarelli et al.~\citeyear{ceccarelli14}). {\it Left panel:} solution
  for FIR4-\CV\ and FIR5, where the the measured \CIII/\CV\ abundance
  ratio is 4--12 and $\leq6$ respectively. The  cosmic-ray ionisation rate is $4\times
  10^{-14}$ s$^{-1}$ and the time is $3\times10^4$ yr.  {\it Right
  panel:} a possible solution  for FIR4-low\CV, where the 
  measured \CIII/\CV\ abundance ratio is 10--30. The  cosmic-ray
  ionisation rate is $1\times 10^{-17}$ s$^{-1}$ and the time is $1\times10^5$ yr.}
\end{center}
\end{figure*}

In order to find the solution(s) that best fit the observations, we
plotted the contours of the HC$_3$N/HC$_5$N abundance ratio and
overlapped them with the HC$_3$N abundance contours. For the latter,
we considered the range 0.5--5.0$\times10^{-11}$ (see Sect.~\ref{sec:deriv-ciiicv-abund}). 
We produced plots at $1\times10^4$,
$3\times10^4$, $1\times10^5$, $3\times10^5$, and $1\times10^6$ yr.
Sample results discussing the case of $\zeta = 4\times 10^{-14}$ s$^{-1}$
and $\zeta = 1\times 10^{-17}$ s$^{-1}$ are shown in Fig. \ref{fig_ratios}.
All plots produced for the different ionisation rates at five 
different times from $1\times 10^4$ yrs to $1\times 10^6$ yrs
are shown in Appendix~\ref{appd}.
Note that the gaseous C/O abundance ratio is
very likely to be lower than 1 because, when the oxygen starts to be
trapped in water ices, CO is already formed and, therefore, has
already segregated as much oxygen as carbon. Therefore, we marked the
C/O$\geq$1 region in the figure as unlikely.

\noindent
{\it FIR4-\CV:}
We first discuss the case of FIR4-\CV, where the measured \CIII/\CV\
abundance ratio is 4--12.  We found no solutions that simultaneously
reproduce the measured HC$_3$N abundance and HC$_3$N/HC$_5$N abundance
ratio, if $\zeta= 1\times10^{-17}$ or $3\times10^{-16}$ s$^{-1}$. On
the contrary, we found solutions when $\zeta= 4\times10^{-14}$
s$^{-1}$ is considered. Specifically, the best solution, shown in the
left panel of Fig. \ref{fig_ratios}, is found at $3\times10^4$ yr and
provides stringent constraints on the gaseous elemental abundances of
carbon and oxygen as well: the C/O abundance ratio must be very close
to unity, O/H has to be lower than $1.4\times10^{-4}$ and C/H lower
than $1.3\times10^{-4}$.  Slightly different ($\sim 20 \%$) solutions
are also found for an age within $1\times10^4$ and $3\times10^5$ yr.

\noindent
{\it FIR4-low\CV:}
In FIR4-low\CV\ the measured \CIII/\CV\ abundance ratio is 10--30. In
this case, a solution is found for $\zeta= 4\times10^{-14}$ s$^{-1}$
similar to the FIR4-\CV, but with slightly larger C/O ratios, as shown
in the left panel of Fig. \ref{fig_ratios}. However, solutions are
also found with lower $\zeta$. The right panel of
Fig. \ref{fig_ratios} shows the case obtained with
$\zeta= 1\times10^{-17}$ s$^{-1}$ and time$=1\times10^5$ yr. An
\CIII/\CV\ abundance ratio between about 15 and 30 can be reproduced with a
carbon abundance between 9 and 6 $\times10^{-5}$.  Note that
no solutions exist for $< 1\times10^5$ and $\> 3\times10^5$ yr.

\noindent
{\it FIR5:}
In FIR5 the measured \CIII/\CV\ abundance ratio is lower than
6. Assuming a gas temperature and density similar to those in FIR4, 
the left panel of Fig. \ref{fig_ratios}  applies to this source too. No
solutions are found for $\zeta < 4\times10^{-17}$ s$^{-1}$. The
constraint on the C/O gaseous elemental abundance ratio is even more
stringent than in FIR4-\CV, and has to be very close to unity.

\section{Discussion and conclusions}
\label{discu}

The high-spatial resolution SOLIS observations show that cyanopolyynes
are present in the three sources encompassed by the NOEMA primary beam:
FIR3, FIR4 and FIR5. In FIR3, only \CIII\ is detected, whereas both
\CIII\ and \CV\ are detected in FIR4 and FIR5. FIR4 is composed of two
regions: an Eastern half, FIR4-\CV, in which the \CV\ is strong, and a
Western half, FIR4-low\CV , with faint \CV\ emission. The measured
\CIII/\CV\ abundance ratio is $\le6$, 4--12 and 10--30 in FIR5,
FIR4-\CV\ and FIR4-low\CV , respectively.

Our analysis indicates that the FIR4-\CV\ and FIR5 sources are young,
with an age between $10^4$ and $3\times 10^5$ yr, and demonstrates the
presence of a high cosmic-ray ionisation rate,
$\zeta \sim 4\times10^{-14}$ s$^{-1}$, permeating these sources. In
addition, the C/O gaseous elemental abundance ratio is very close to
unity, the O abundance is $\leq 1.4\times10^{-4}$ and the carbon
abundance is $\leq 1.3\times10^{-4}$. Conversely, the
lower value of the \CIII/\CV\ abundance ratio in FIR4-low\CV\ can
either be due to a lower cosmic rays ionisation rate or to a larger
C/O gaseous abundance ratio.

The C/O gaseous elemental abundance ratio being close to unity, and the
relatively low carbon and oxygen abundances, are consistent with the
idea that FIR4 and FIR5 have passed through a cold phase where icey
mantles have locked up a large fraction of oxygen (probably in the
form of H$_2$O) and carbon (e.g.~iced CO, CO$_2$, CH$_4$; e.g. Boogert
et al.~\citeyear{boogert}, Ruffle et al.~\citeyear{ruffle}). 
If the hypothesis that the difference between FIR4-\CV\
and FIR4-low\CV\ is caused by a lower C/O ratio in FIR4-low\CV\ is correct,
it would imply that the freezing has been less efficient in
FIR4-low\CV . This could be due to warmer dust, but the source of the
heating is not obvious, or to a lower density, so that the timescale
for freezing is larger. Lacking measurements able to provide the
gradient in density in FIR4, we cannot reach a firm conclusion on
that.

More interesting, the low ($\leq$12) measured \CIII/\CV\ abundance
ratio in FIR5 and FIR4-\CV\ requires an enhanced cosmic-ray ionisation
rate, which is what Ceccarelli et al.~(\citeyear{ceccarelli14}) claimed for FIR4 based on
a totally different dataset (high-J HCO$^+$ and N$_2$H$^+$ lines
observed by Herschel). Ceccarelli et al. argued that the enhanced
cosmic-ray ionisation rate is due to an embedded source of high energy
($\geq10$ MeV) particles, similar to what happened to the young
Solar System (see Sect.~1). In addition to confirming that 
claim, these new observations suggest that there maybe a gradient in
the cosmic-ray ionisation rate across the FIR4 condensation, with the
eastern half strongly irradiated and the western region partially
shielded and with a lower $\zeta$. 
Let us now discuss this possibility.  The observed gradient would
be consistent with the observed region of bright free-free emission
partially overlapping with the FIR4-\CV\ region and extending outside the 
eastern border of FIR4 (Reipurth et al.~\citeyear{reipurth}; Lopez-Sepulcre et
al.~\citeyear{lopez13} and discussion there) and the possibility that the source(s)
of energetic particles permeating FIR4-\CV\ and FIR5 lie there. In
addition, there would not be the need to claim a gradient in the
density, the cause of which would be mysterious, and, most importantly, a
very narrow range of possible ages, around $1\times10^5$ yr. We conclude that,
although not fully proven, the hypothesis of one or more sources of
high energy particles to the East of FIR4 and FIR5 remains the most
appealing and consistent to explain the observations so far.

An alternative explanation is a temperature gradient (see
  Sect.~\ref{model}), with temperature decreasing across FIR4 from
  East to West, which, as discussed in Sect.~\ref{model}, would rule out a 
  gradient in the irradation by cosmic rays. However, for gradients of 
  10--20~K, a \CIII /\CV \ abundance ratio $\leq 14$ can be reproduced
  only if a high cosmic ray ionisation rate is present. Therefore,
  this possibility is still consistent with a high irradiation of
  FIR4.

In conclusion, even though the \CIII /\CV\ line intensity ratio gradient in 
FIR4 can be caused by a gradient in temperature, the measured \CIII /\CV\ 
abundance ratio points, anyway, to a strong irradiation by cosmic-rays/energetic 
particles of the region. We find that energetic particle irradiation promotes the 
production of carbon chains. As irradiation was also present during the early 
phases of our Solar System, it is tempting to speculate that such energetic 
processes have also promoted the production of important carbon reservoirs 
in the Solar Nebula. Such reservoirs could then be delivered to the early Earth 
to foster pre-biotic chemistry evolution.  Whether the energetic particle irradiation 
experienced by the young Solar Nebula and the emergence of life on Earth are 
connected is a wild, but intriguing hypothesis which cries for more dedicate work, 
possibly in connection with biophysicists.

{\it Acknowledgments.}  We thank the IRAM staff for its help in the
data reduction. We also thank the anonymous Referee for his/her
constructive comments. This work was supported by the French program
Physique et Chimie du Milieu Interstellaire (PCMI) funded by the
Conseil National de la Recherche Scientifique (CNRS) and Centre
National d'Etudes Spatiales (CNES), and by a grant from LabeX 
Osug@2020 (Investissements d’avenir - ANR10LABX56). 
Partial salary support for A. Pon was provided by a Canadian 
Institute for Theoretical Astrophysics
(CITA) National Fellowship. PC, A.Punanova, AC, and JEP acknowledge support from the
European Research Council (project PALs 320620).
CF acknowledges founding from French space agency CNES.

{}

\clearpage

\renewcommand{\thetable}{A-\arabic{table}}
\renewcommand{\thefigure}{A-\arabic{figure}}
\renewcommand{\thesection}{A-\arabic{section}}
\setcounter{section}{0}
\setcounter{table}{0}
\setcounter{figure}{0}
\section*{Appendix A}
\label{appa}

\section{Derivation of missing flux}
\label{appa_missingflux}

\begin{figure}
\begin{center}
\includegraphics[angle=0,width=9cm]{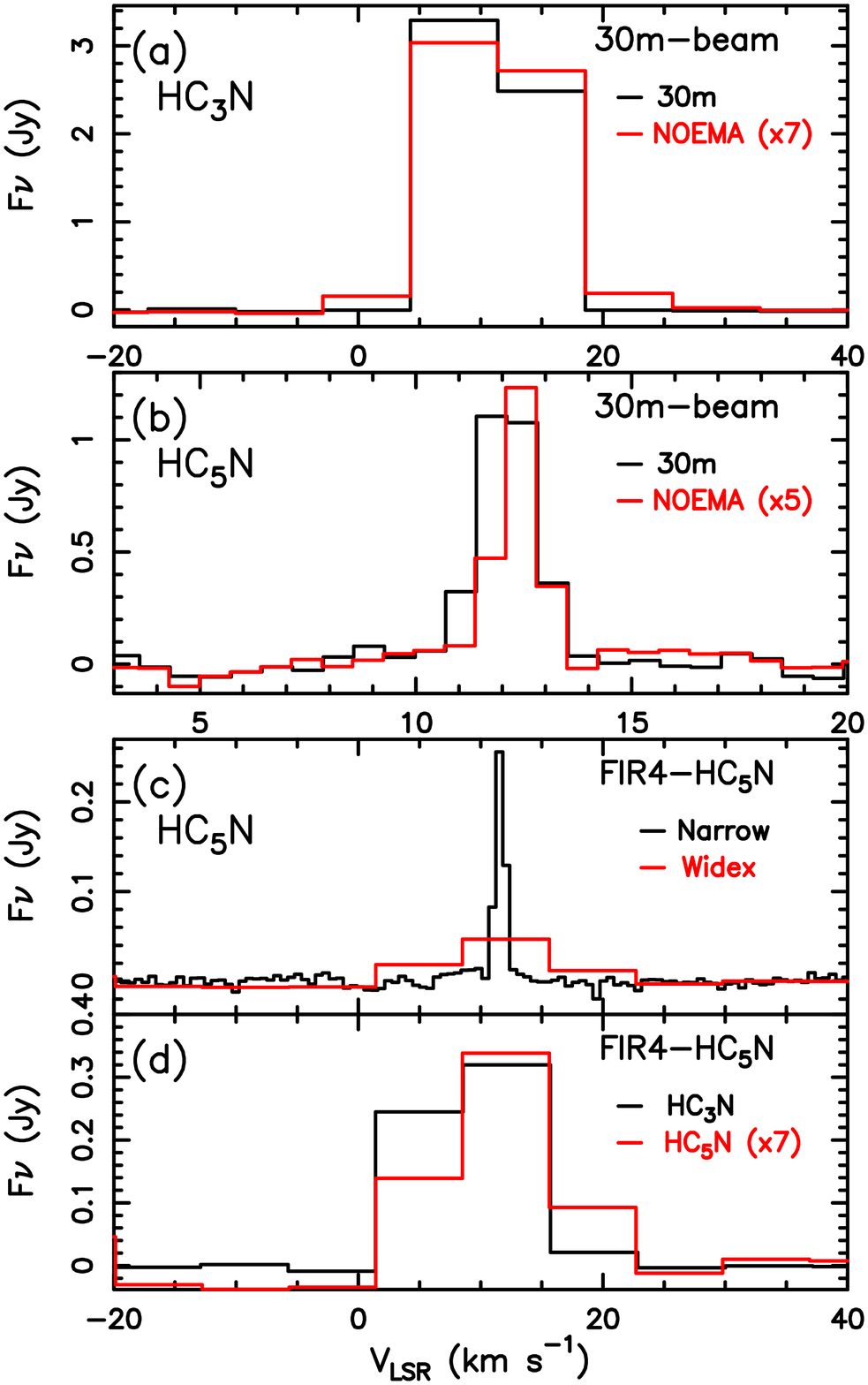}
\caption[]
{\label{flux_comp}{{\it (a):} spectrum of \CIII\ (9--8) obtained with 
the IRAM-30m Telescope (black histogram) in the framework of the ASAI
large program, and SOLIS-NOEMA spectrum (red histogram) extracted from
a circular region equal to the IRAM-30m HPBW ($\sim $30\asec ).
Note that the flux resulting from the convolved NOEMA observations have 
been multiplied by a factor 7 to match those of the 30m Telescope:
we thus recover about the $15\%$ of the total flux. 
\newline
{\it (b):} same as panel {\it (a)} for \CV\ (31--30).
In this case, the NOEMA spectrum has been multiplied by a factor 5.
From the plot, it is apparent that we recover slightly less than the $20\%$ 
of the total flux. 
\newline
{\it (c):} comparison between the Narrow (black) and Widex (red)
spectrum of \CV\ (31--30), both integrated on the region FIR4-\CV\
(red contour in Fig.~\ref{fig_maps}). The integral under
the channel with signal is different by $\sim 7\%$ (see Sect.~\ref{obs}).
\newline
{\it (d):} Widex spectra of \CIII\ (black) and \CV\ (red) intergated over
the same region as in panel {\it (c)}. The \CV\ spectrum has been multiplied
by 7 for clarity of the figure.}}
\end{center}
\end{figure}

To estimate how much flux is resolved out by the interferometer, 
we have compared IRAM-30m spectra with interferometric spectra extracted 
from a region corresponding to the beam of the single-dish observations
($\sim 30$\asec ). 
The IRAM-30m spectra have been converted from main beam temperature units 
($T_{{\rm MB}}$) to flux density units ($F_{\nu}$) by assuming that the telescope
beam is Gaussian and the source size is smaller than the beam, via the formula 
$F_{\nu}=2 k (\nu^2/c^2 ) [(\pi /4\ln 2)(\Theta_{\rm MB})^2]\; T_{{\rm MB}}$,
where $k$ is the Boltzmann constant, $\Theta_{\rm MB}$ is the half
power beam width of the IRAM-30m Telescope, and $\nu$ is the line 
rest frequency. The resulting spectra are shown in panels {\it (a)} and {\it (b)}
of Fig.~\ref{flux_comp}: 
with the NOEMA we are recovering $\sim 14-15\%$ of the flux detected 
with the IRAM-30m telescope in \CIII , and $\sim 20\%$ in \CIII .

\section{IRAM-30m spectra of $^{13}$C isotopologues of \CIII }
\label{appc}

\begin{figure*}
\begin{center}
\includegraphics[angle=0,width=17cm]{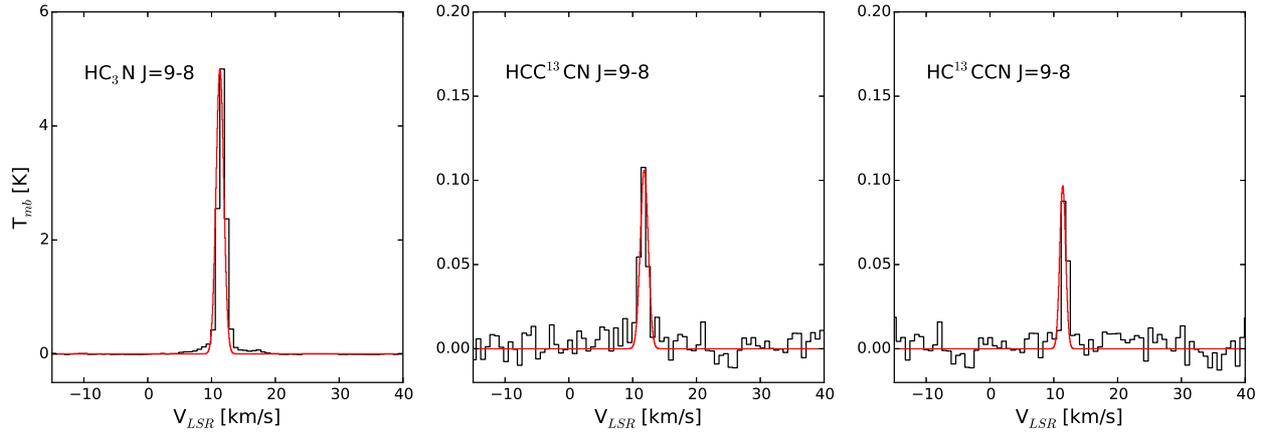}
\caption[]
{\label{hc3n-iso}{IRAM-30m spectra of the \CIII , HCC$^{13}$CN and HC$^{13}$CCN
(9--8) line (Jaber Al-Edhari et al., in preparation). The (9--8) line of the third isotopologue, 
H$^{13}$CCCN, lies outside the observed frequency range.}}
\end{center}
\end{figure*}

In Fig.~\ref{hc3n-iso}, we show the IRAM-30m spectrum of \CIII\ (9--8), and those 
of the two isotopologues HCC$^{13}$CN and HC$^{13}$CCN in the same transition.
The integrated intensities for the three transitions are: $7.1\pm 0.7$ K \kms\ in \CIII ,
 $0.16\pm 0.02$ K \kms\ in HCC$^{13}$CN, and $0.11\pm 0.01$ K \kms\ in HC$^{13}$CCN. 
Please note that the (9--8) line of the third isotopologue, H$^{13}$CCCN, lies outside 
the observed frequency range. Using the other two detected isotopologues, the $^{12}$C/$^{13}$C 
line ratio is $44\pm 8$ and $64\pm 12$, respectively. Therefore, considering an elemental 
$^{12}$C/$^{13}$C ratio of $\sim 68$ for the local interstellar medium (Milan et 
al.~\citeyear{milam2005}), the line of the main isotopologue is optically thin or at most 
slightly thick. Note that the difference in the $^{12}$C/$^{13}$C intensity ratio 
between the two isotopologues is present in all the other detected lines and it is, 
therefore, real and due to the \CIII\ chemistry, specifically its formation route. 
This is the focus of a forthcoming article by Jaber Al-Edhari et al. (in preparation).

\section{New reactions rates for HC$_5$N}
\label{appb}

We adopted as a base the KIDA chemical database for the network of
reactions. We modified it following the works by Loison et al. (\citeyear{loison}),
Balucani et al. (\citeyear{balucani}) and Barone et al. (\citeyear{barone}). In addition, we
modified the rate $k$ of two reactions involving the formation of HC$_5$N and
included a new reaction, following the experimental results from Cheikh (\citeyear{cheikh}), 
Fournier (\citeyear{fournier14}) and Fournier et al. (in prep.). The new rates are listed 
in Tab.~\ref{tab:newreac}, where we reported the coefficients $\alpha$, $\beta$ and 
$\gamma$ defined, as usual, by the formula:

\begin{equation}
\label{eq:1}
 k = \alpha \frac{T}{300K}^\beta exp[-\gamma/T]
\end{equation}

\begin{table}
  \centering
  \begin{tabular}{lcccl}
\hline
    Reaction & $\alpha$ & $\beta$ & $\gamma$ & Ref. \\
\hline
CN + C$_4$H$_2$ $\rightarrow$ HC$_5$N + H   & $4.06 \times 10^{-10}$ &-0.24 & 11.5 & 1, 3\\
C$_2$H + HC$_3$N $\rightarrow$ HC$_5$N + H & $3.91 \times 10^{-10}$ &-1.04 & 0 & 1, 3\\
C$_3$N + C$_2$H$_2$ $\rightarrow$ HC$_5$N + H & $3.09 \times 10^{-10}$ &-0.58 & 33.6 & 2, 3\\
\hline
\end{tabular}
\caption{Modified and new reactions involving the formation of HC$_5$N. References: 1- Cheikh (2012); 2- Fournier (2014); 3- Fournier et al. (2017).}
  \label{tab:newreac}
\end{table}

\section{Chemical model predictions}
\label{appd}

\begin{figure}
\begin{center}
{\includegraphics[angle=0,width=6.2cm]{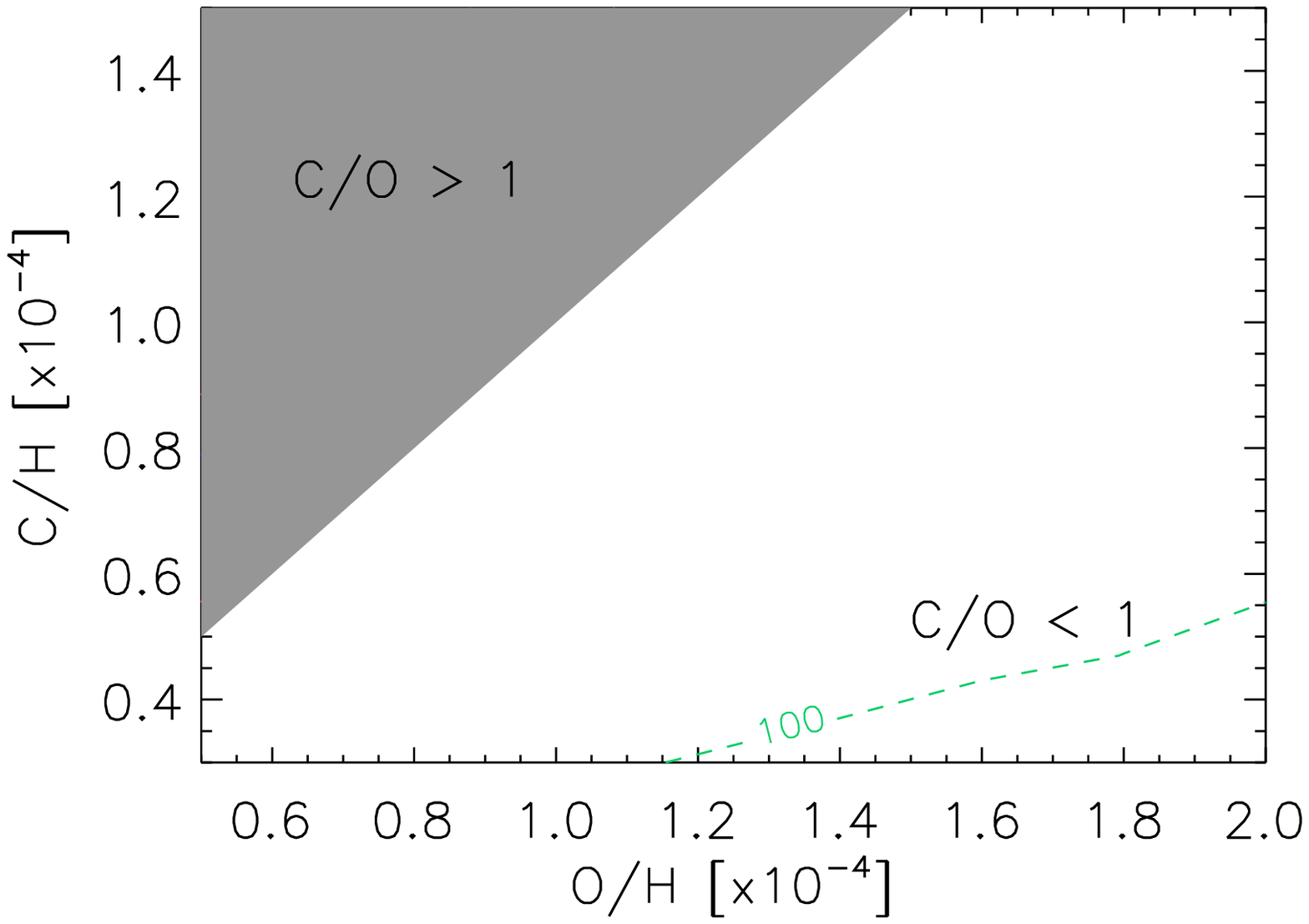}}
{\includegraphics[angle=0,width=6.2cm]{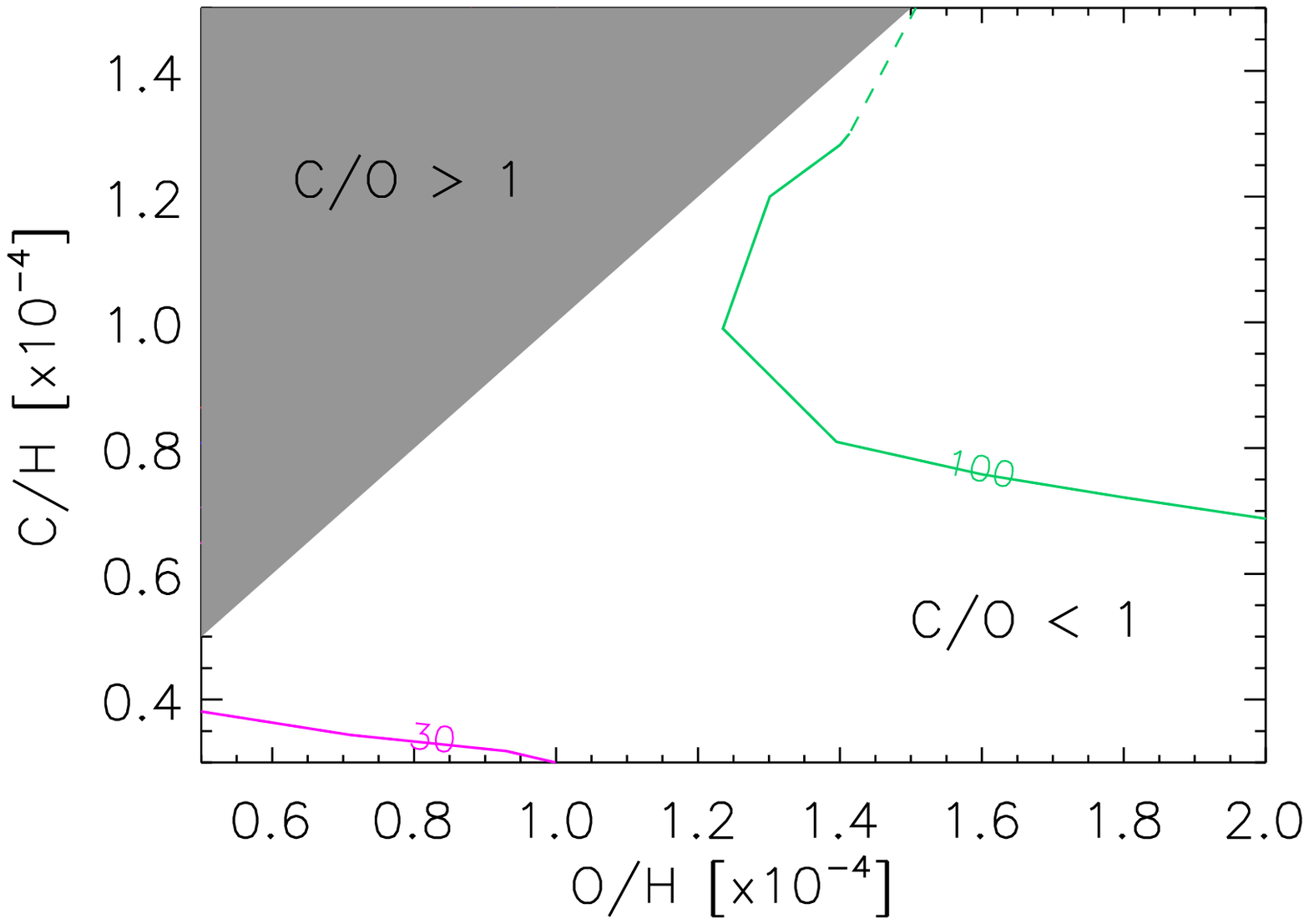}}
{\includegraphics[angle=0,width=6.2cm]{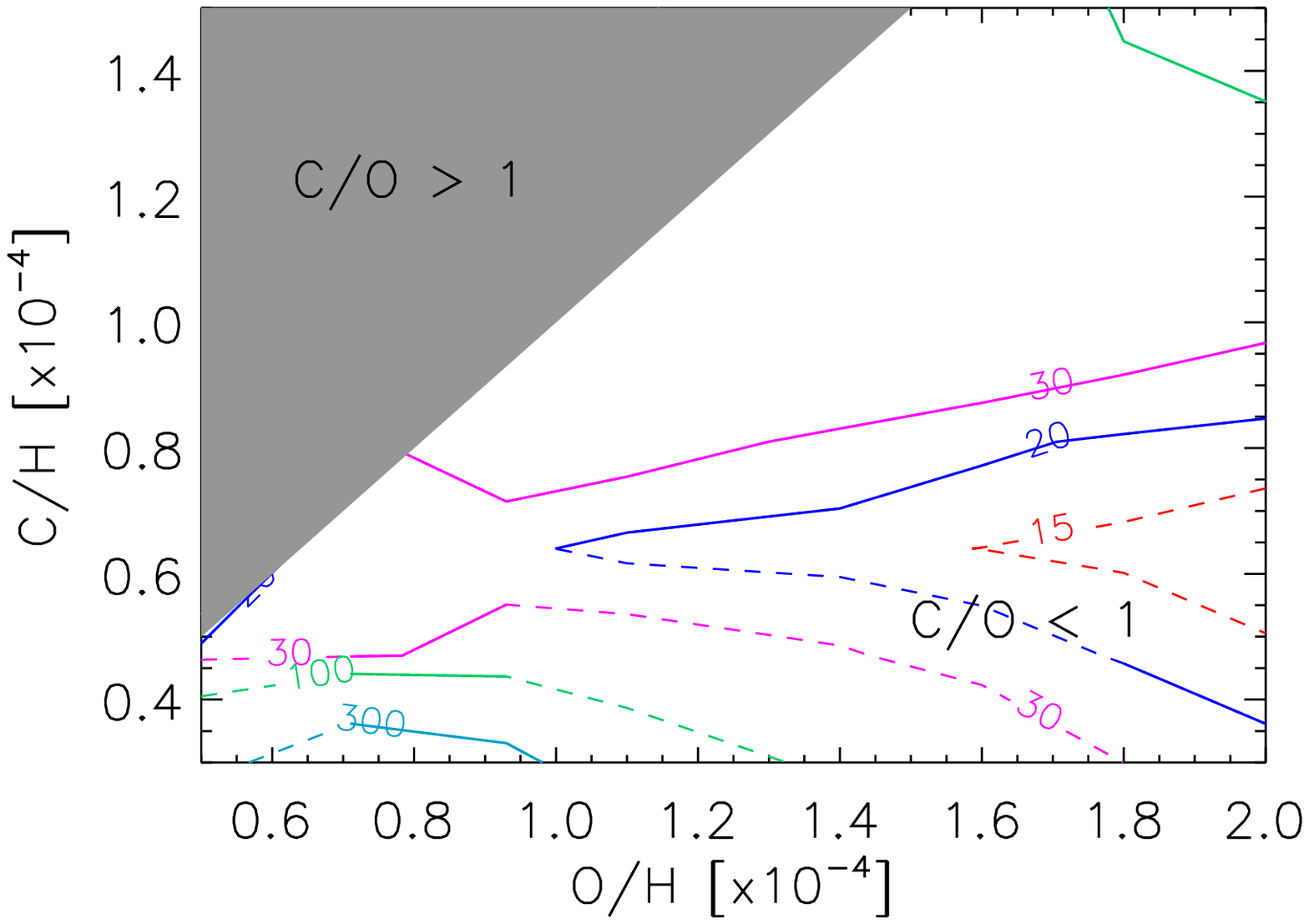}}
{\includegraphics[angle=0,width=6.2cm]{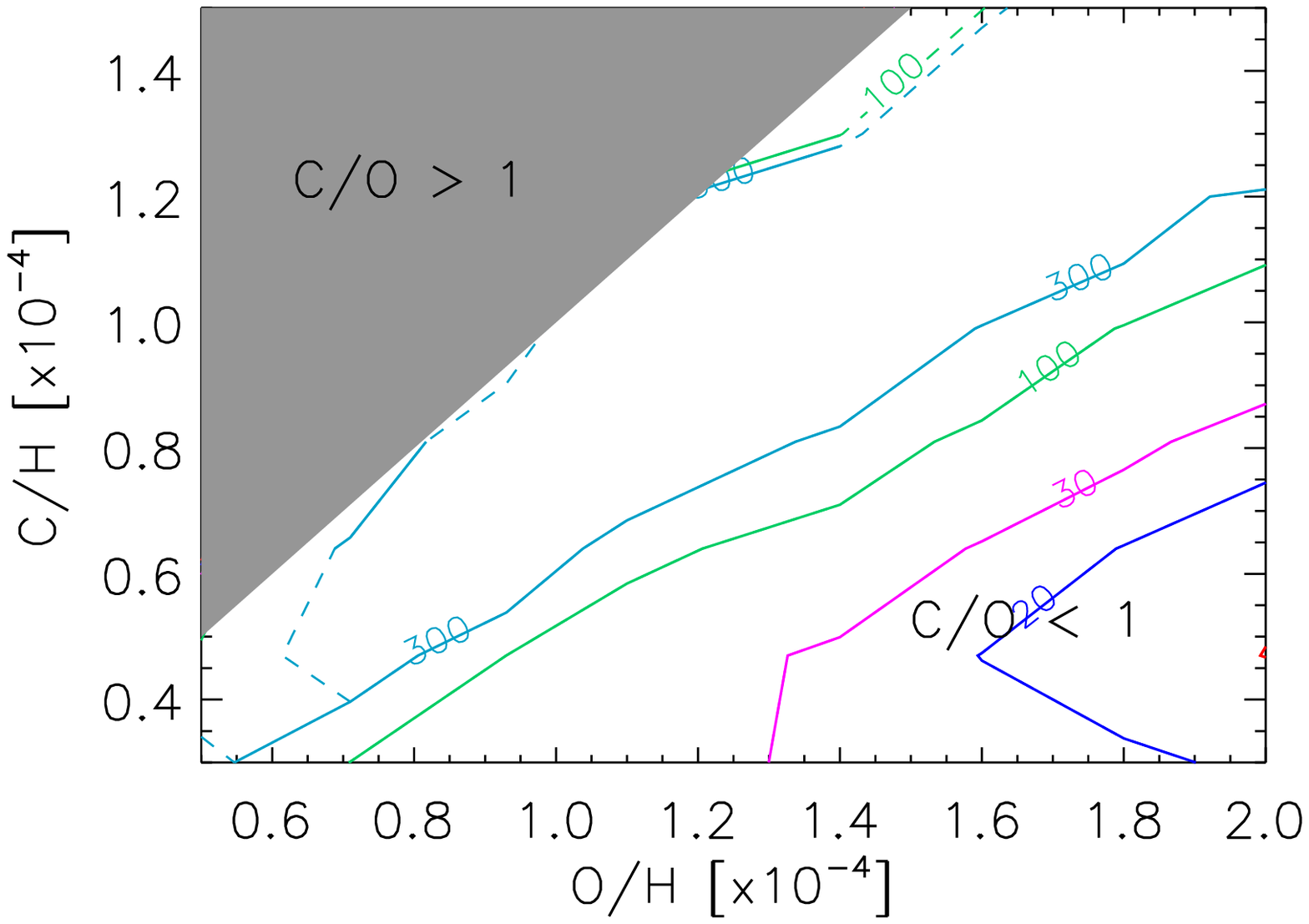}}
{\includegraphics[angle=0,width=6.2cm]{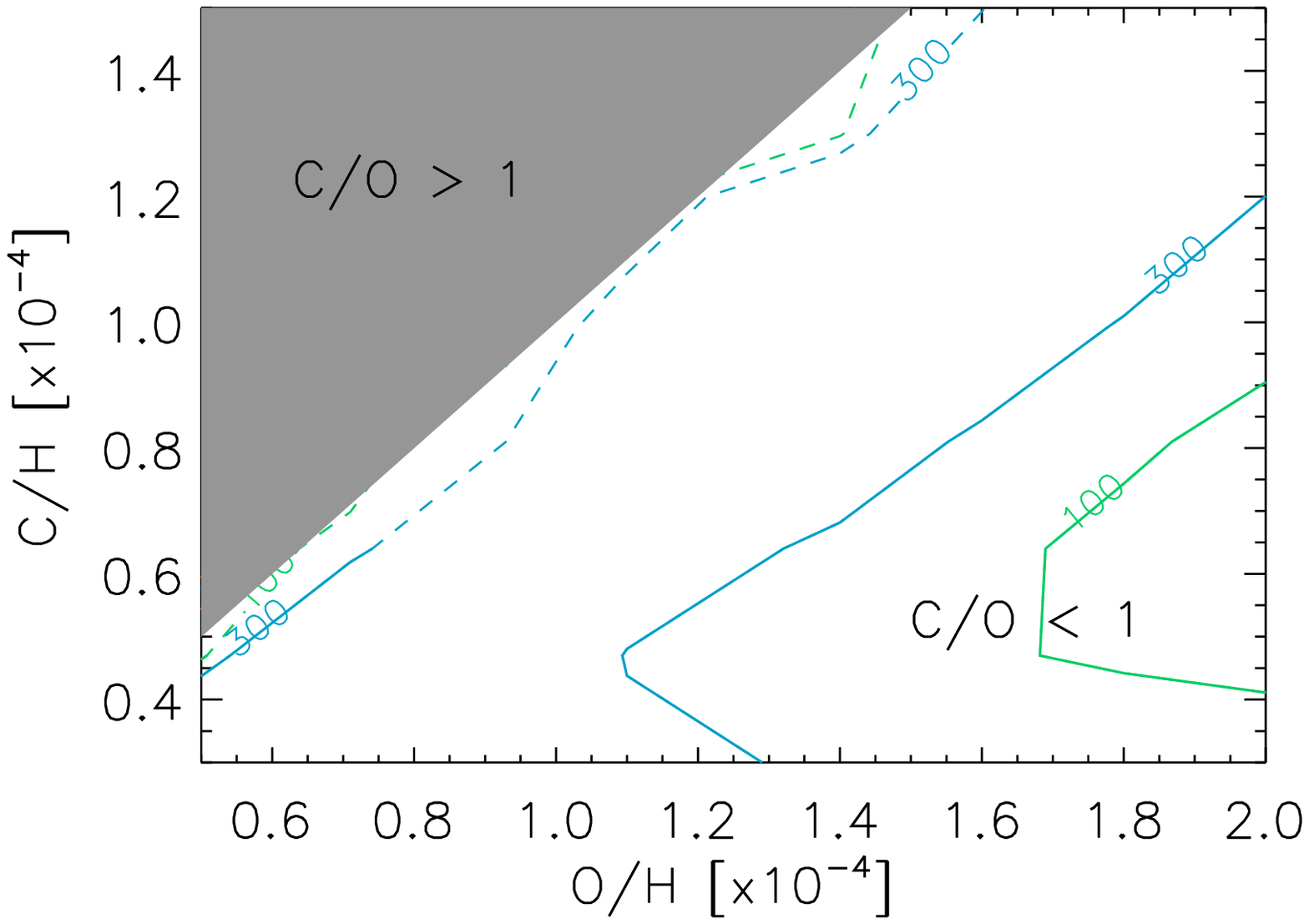}}
\caption[]
{\label{appd1}{Predictions of the \CIII /\CV\ abundance ratio from the chemical models 
described in Sect.~\ref{chemistry} for $\zeta = 1\times 10^{-17}$ at the five different 
times (from top to bottom): $1\times 10^4$ yrs, $3\times 10^4$ yrs, 
$1\times 10^5$ yrs, $3\times 10^5$ yrs, and $1\times 10^6$ yrs. 
All symbols used (contour plots, grey zone, etc.) are the same as in Fig.~\ref{fig_ratios}.}}
\end{center}
\end{figure}

\begin{figure}
\begin{center}
{\includegraphics[angle=0,width=6.2cm]{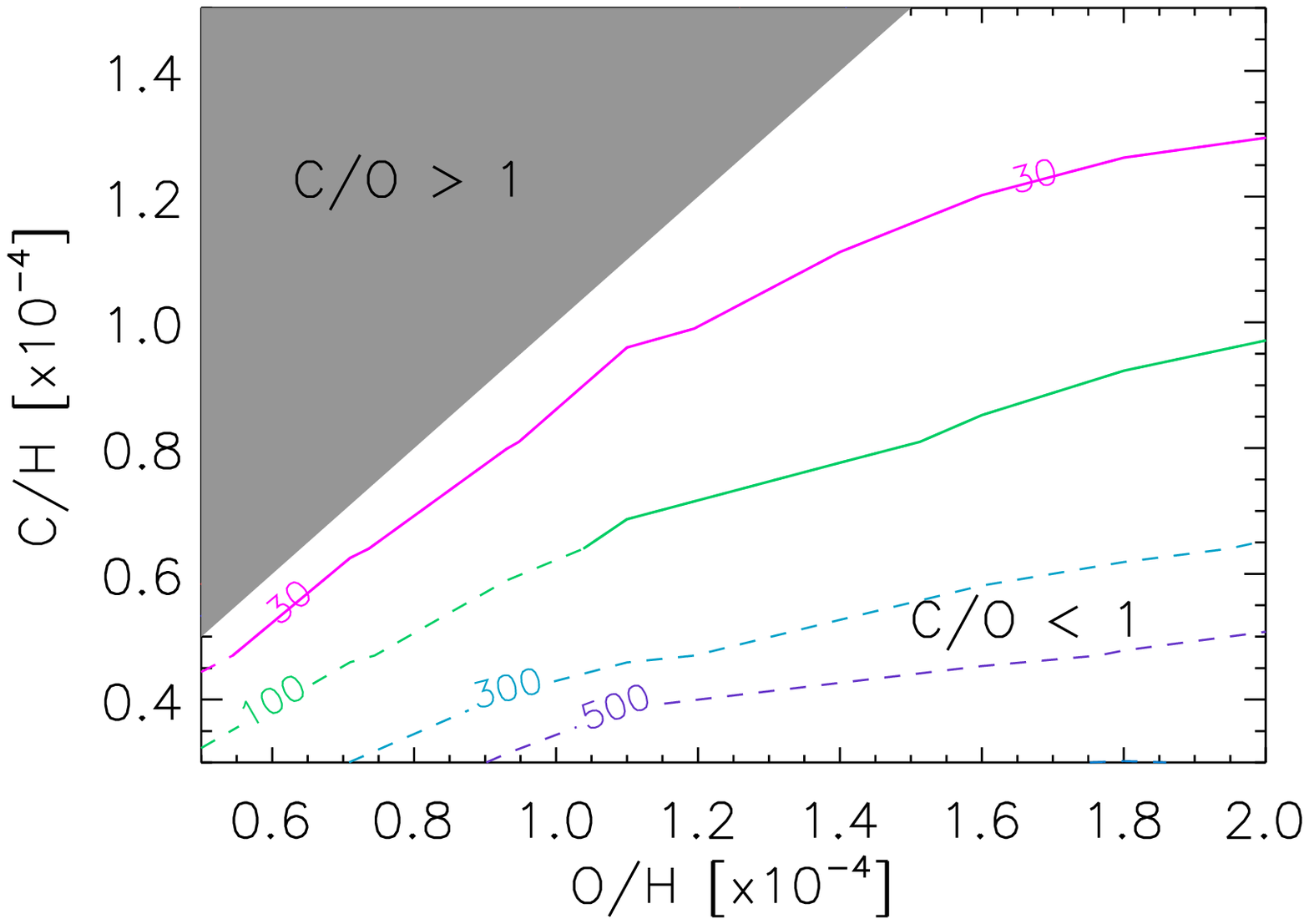}
\includegraphics[angle=0,width=6.2cm]{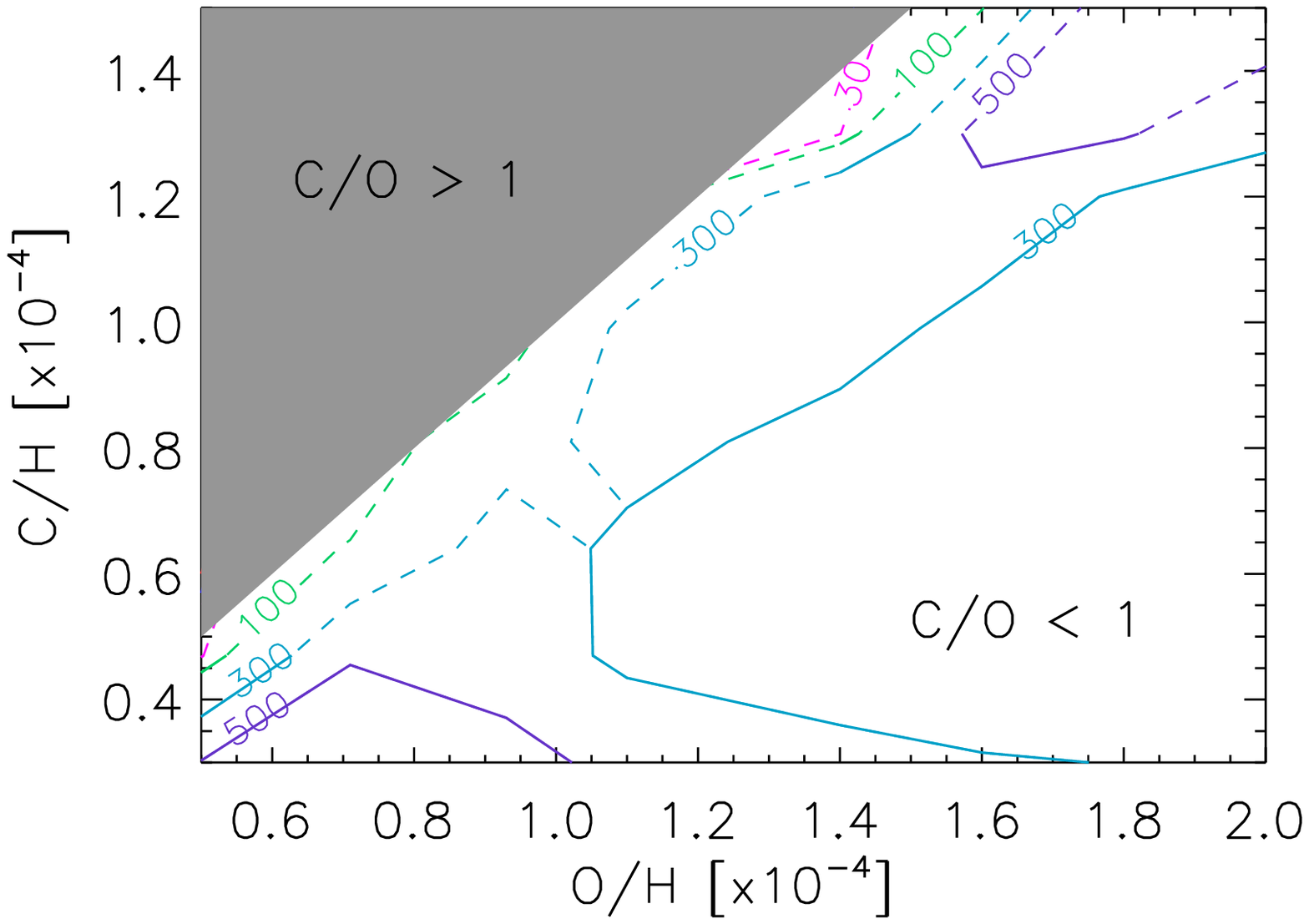}
\includegraphics[angle=0,width=6.2cm]{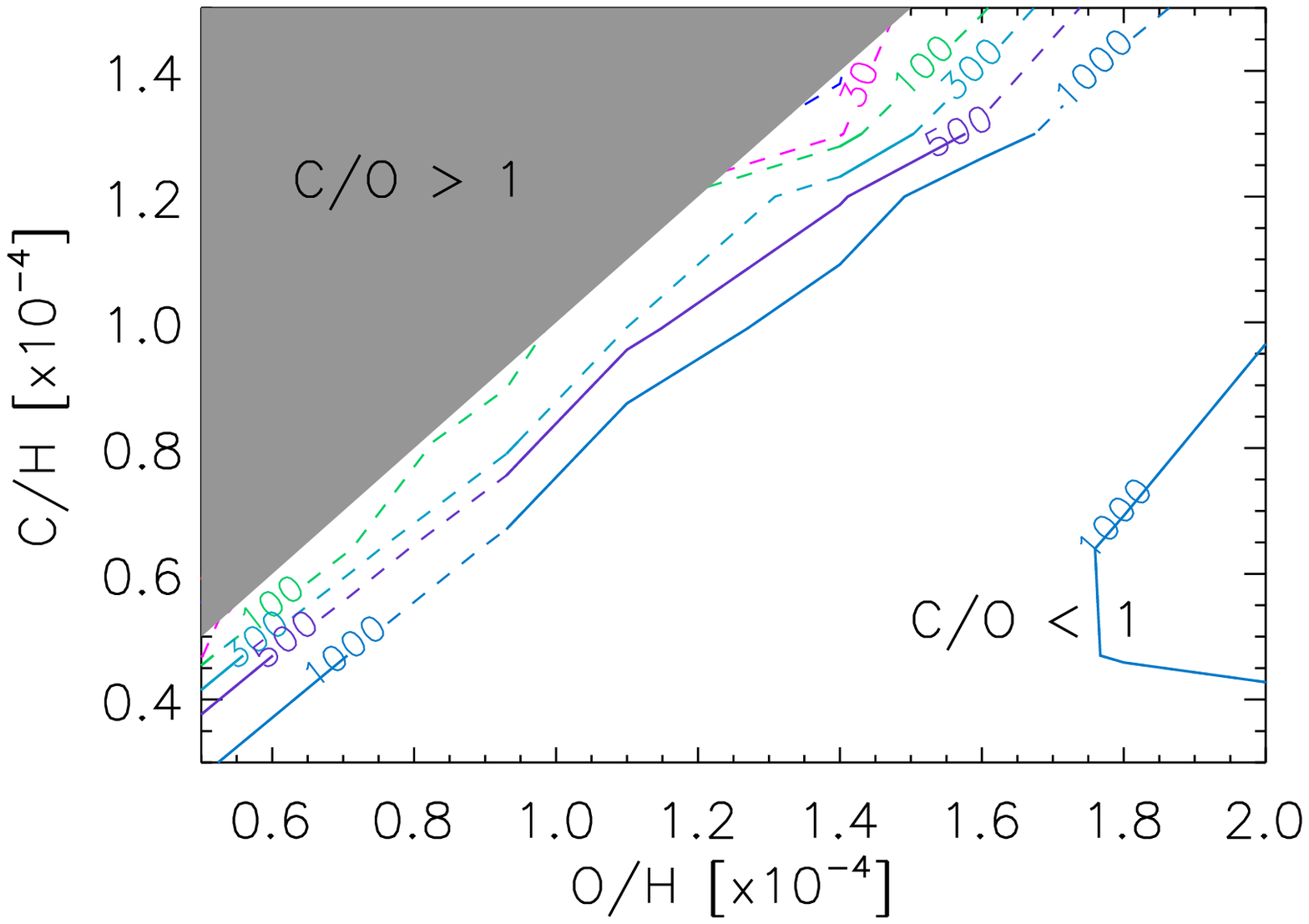}
\includegraphics[angle=0,width=6.2cm]{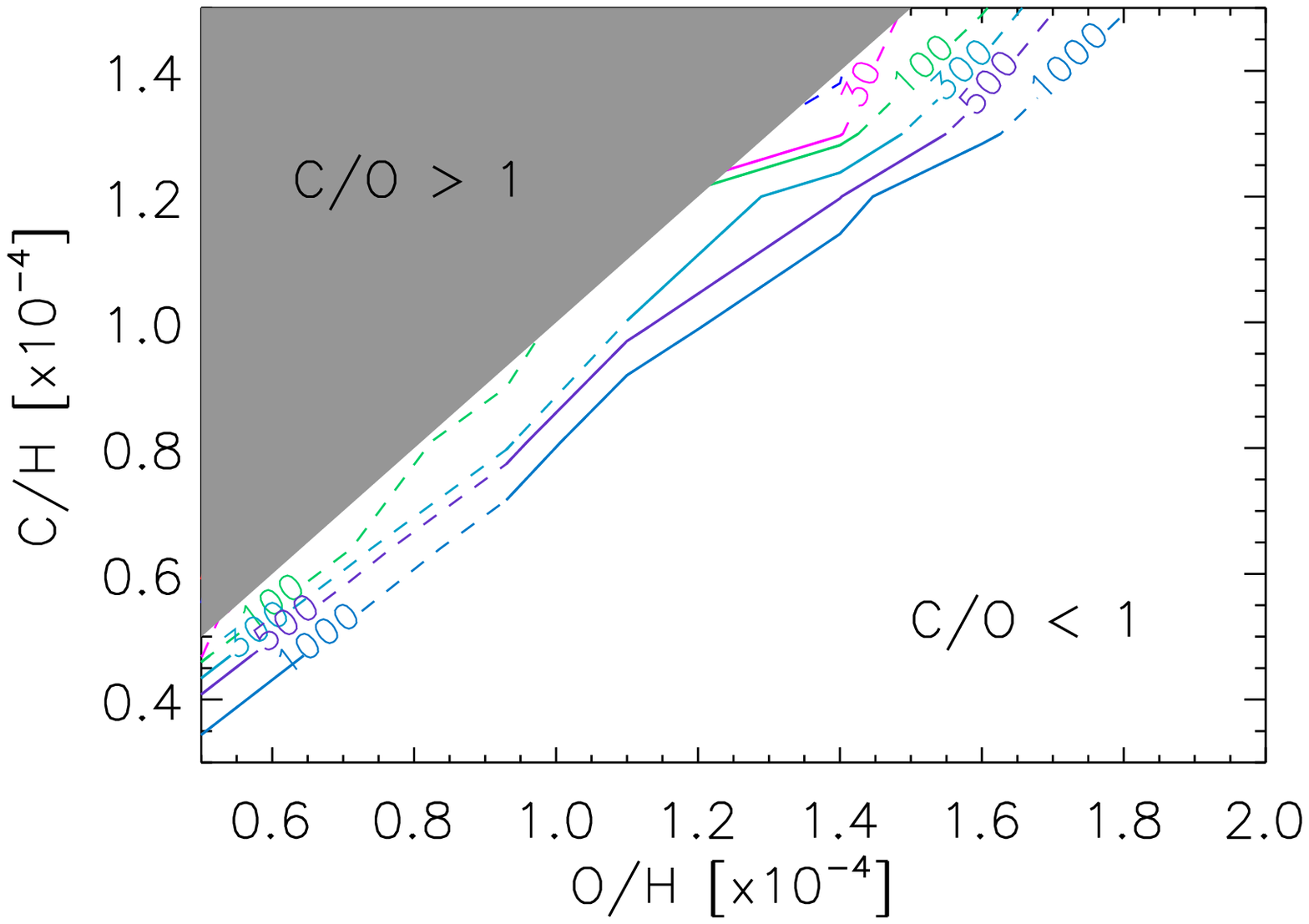}
\includegraphics[angle=0,width=6.2cm]{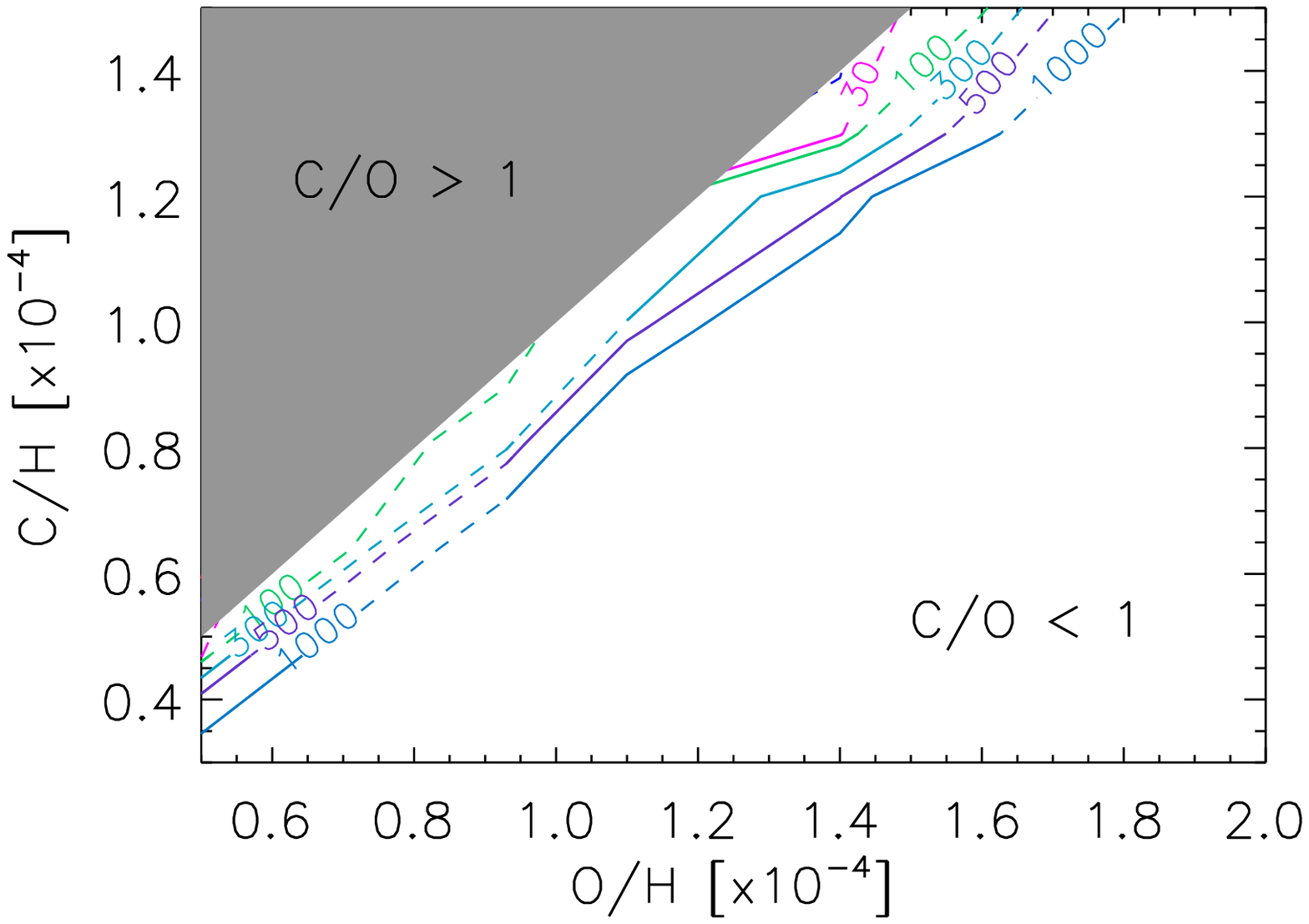}}
\caption[]
{\label{appd2}{Same as Fig.~\ref{appd1} for $\zeta = 3\times 10^{-16}$.}}
\end{center}
\end{figure}

\begin{figure}
\begin{center}
{\includegraphics[angle=0,width=6.2cm]{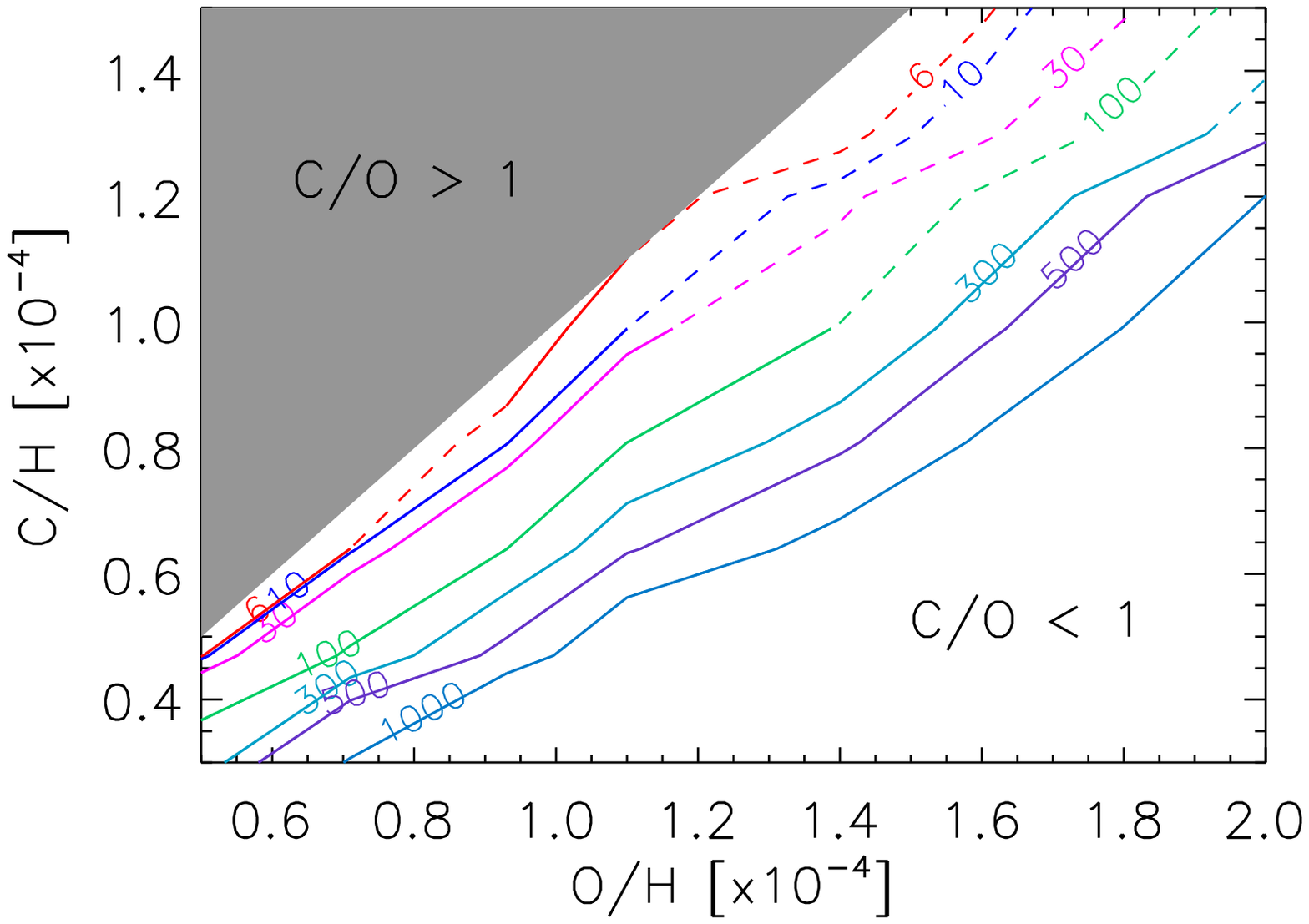}
\includegraphics[angle=0,width=6.2cm]{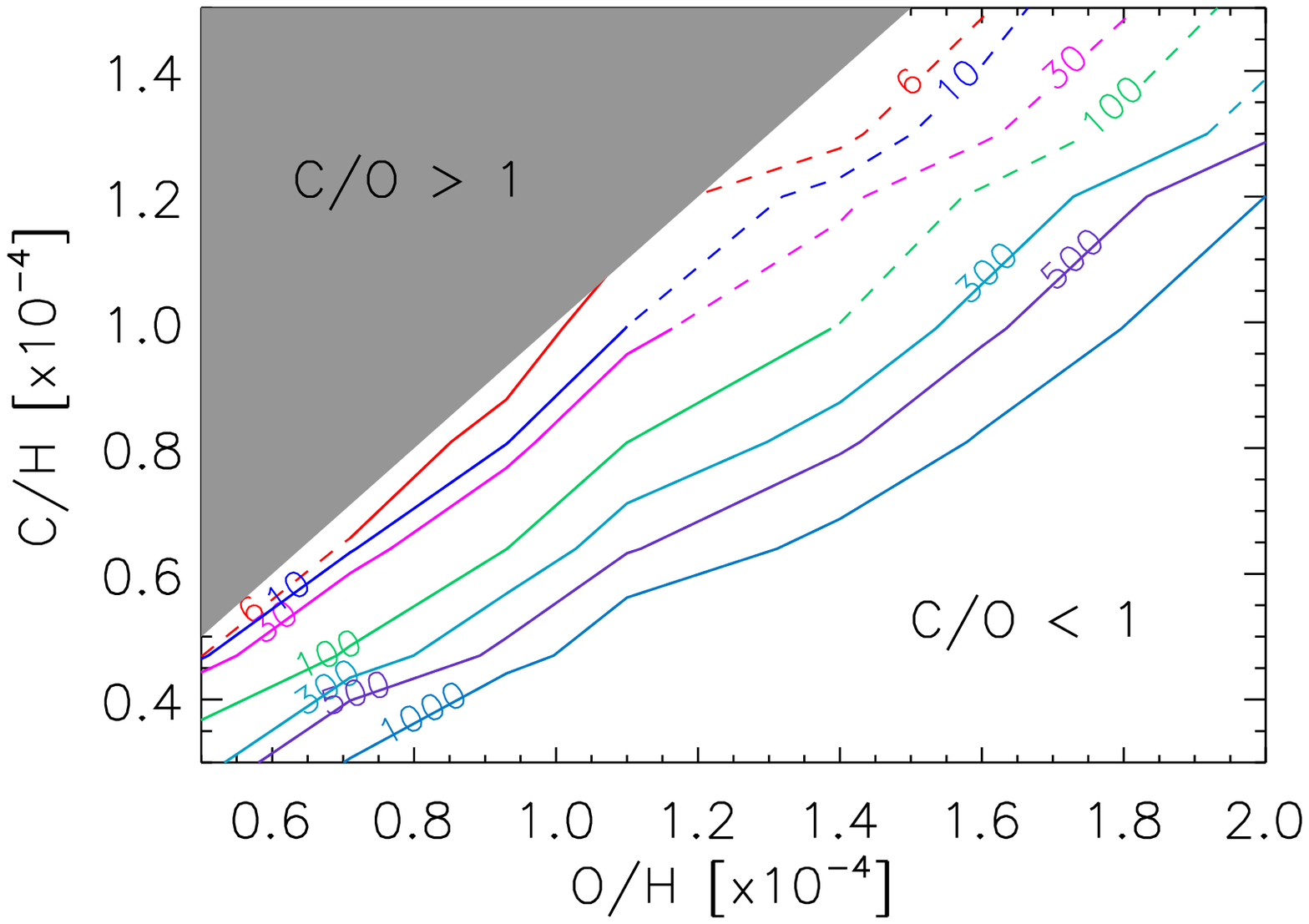}
\includegraphics[angle=0,width=6.2cm]{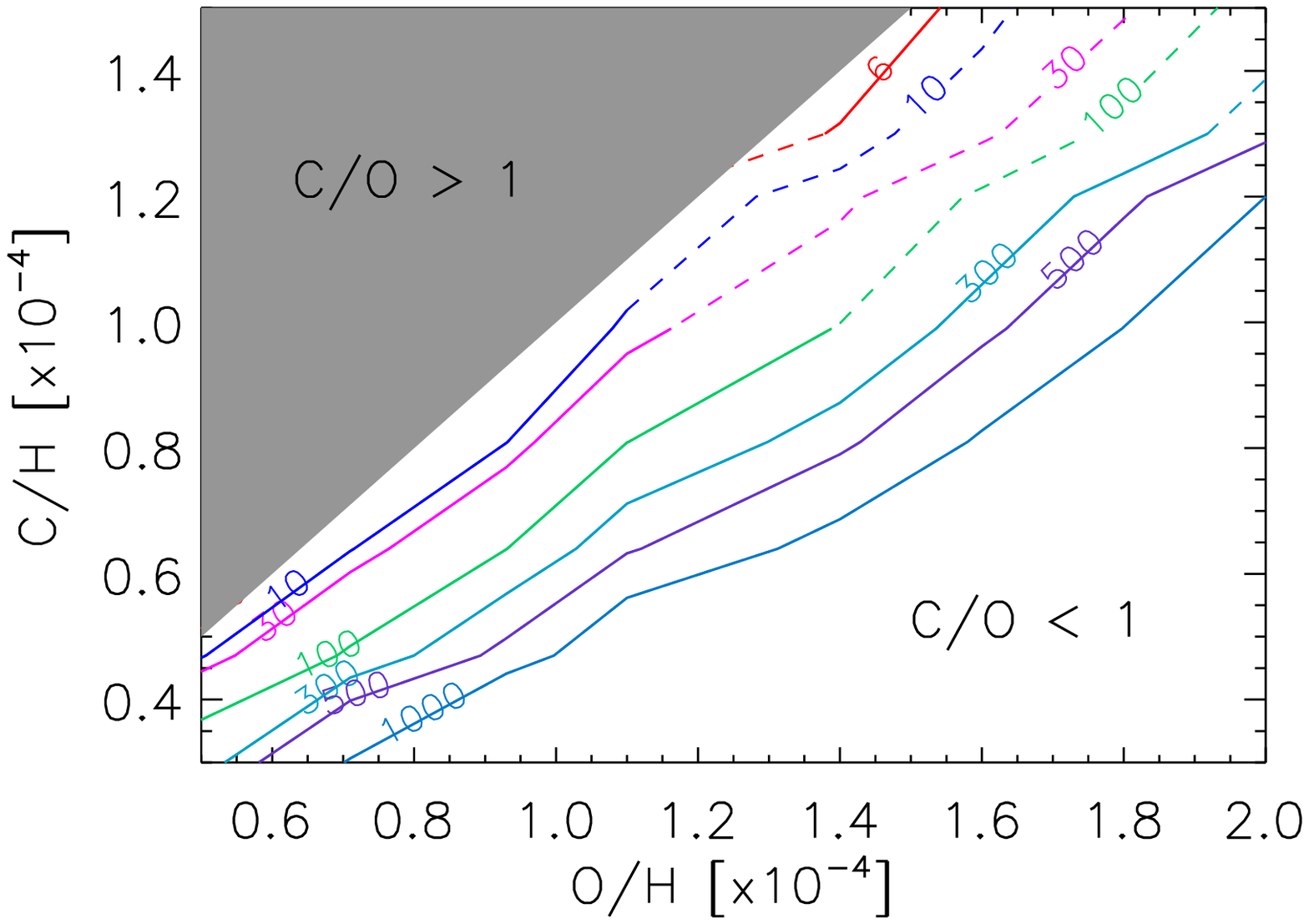}
\includegraphics[angle=0,width=6.2cm]{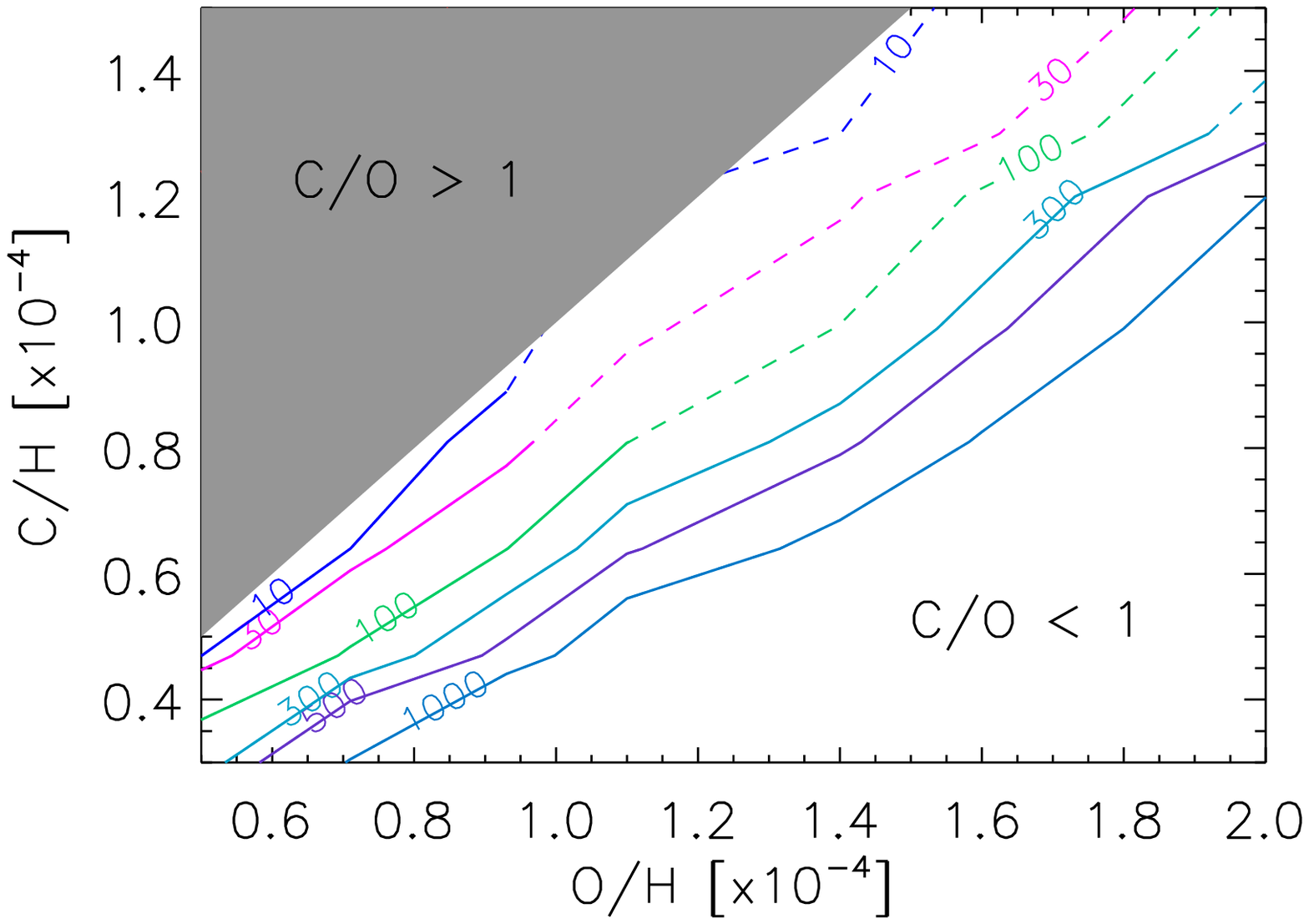}
\includegraphics[angle=0,width=6.2cm]{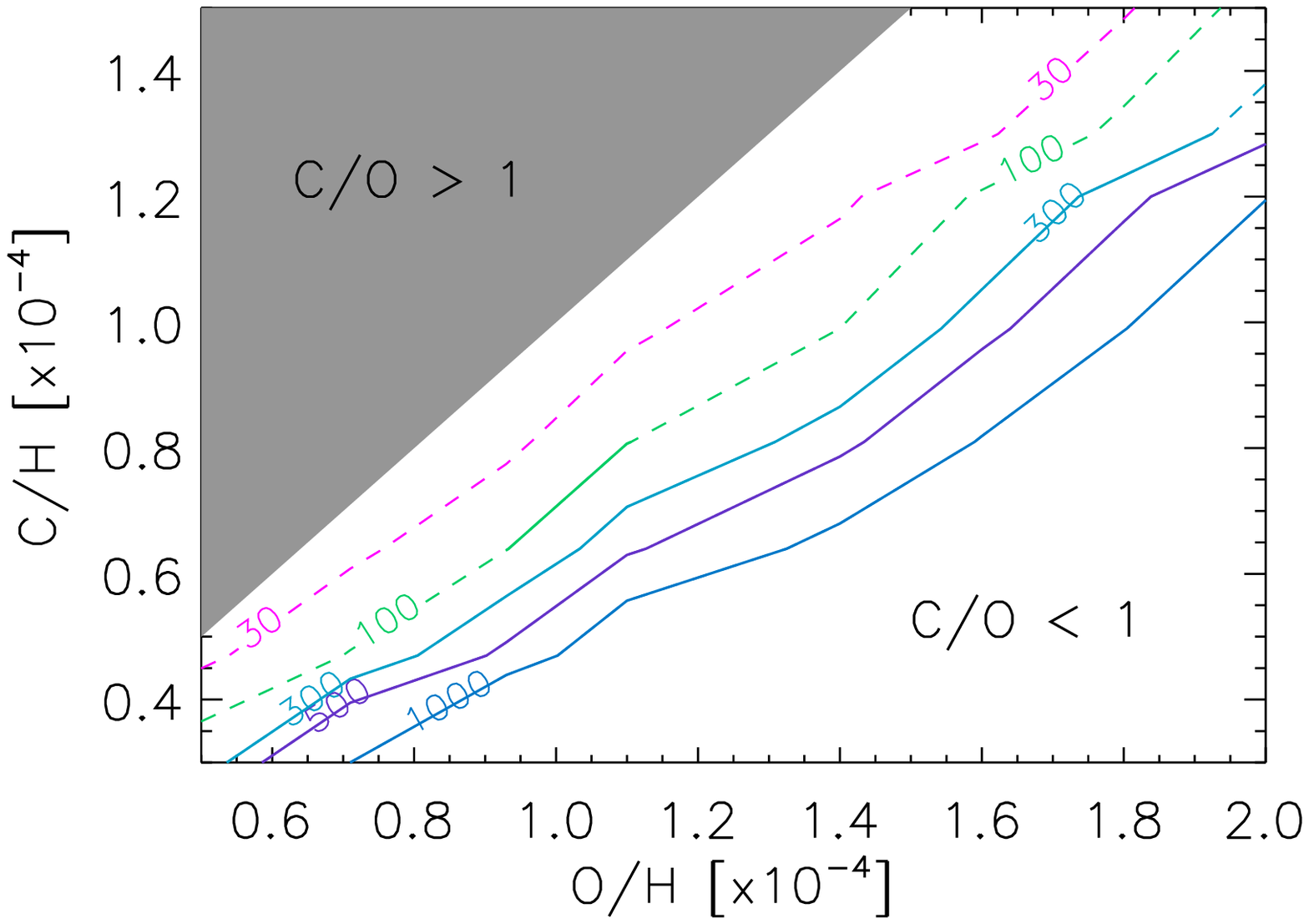}}
\caption[]
{\label{appd3}{Same as Fig.~\ref{appd1} for $\zeta = 4\times 10^{-14}$.}}
\end{center}
\end{figure}

In this Appendix we show the predictions of the chemical model described in 
Sect.~\ref{chemistry} for three different values of the cosmic rays ionisation rate: 
$\zeta = 1\times 10^{-17}$ s$^{-1}$ (Fig.~\ref{appd1}),
$\zeta = 3\times 10^{-16}$ s$^{-1}$ (Fig.~\ref{appd2}), and $\zeta = 4\times 10^{-14}$ s$^{-1}$ 
(Fig.~\ref{appd3}), at five different times: $1\times 10^4$ yrs, $3\times 10^4$ yrs, 
$1\times 10^5$ yrs, $3\times 10^5$ yrs, and $1\times 10^6$ yrs.

%


\begin{thebibliography}{}
\bibitem[Adams(2010)]{adams}
Adams, F.C.~2010, ARA\&A, 48, 47
\bibitem[Adams et al.(2012)]{adams12}
Adams, J.D., Herter, T.L., Osorio, M., Macias, E., Megeath, S.Th., et al.~2012, ApJ, 749, L24
\bibitem[Balucani et al.(2000)]{balucani2000}
Balucani, N., Asvany, O., Huang, L.C.L., Lee, Y.T., Kaiser, R.I.~2000, ApJ, 545, 892
\bibitem[Balucani(2009)]{balucani2009}
Balucani, N.~2009, Int. J. Mol. Sci., 10, 2304
\bibitem[Balucani et al.(2015)]{balucani}
Balucani, N., Ceccarelli, C., Taquet, V.~2015, MNRAS, 449, L16
\bibitem[Barone et al.(2015)]{barone}
Barone, V., Latouche, C., Skouteris, D., Vazart, F., Balucani, N., Ceccarelli, C., Lefloch, B.~2015, MNRAS, 453, L31 
\bibitem[Bell et al.(1992)]{bell92}
Bell, M.B., Avery, L.W., MacLeod, J.M., Matthews, H.E.~1992, ApJ, 400, 551
\bibitem[Bell et al.(1997)]{bell}
Bell, M.B., Feldman, P.A., Travers, M.J., McCarthy, M.C., Gottlieb, C.A., Thaddeus, P.~1997, ApJL, 483, 61
\bibitem[Bisschop et al.(2006)]{bisschop}
Bisschop, S.E., Fraser, H.J., \"{O}berg, K.I., van Dishoeck, E.F., Schlemmer, S.~2006, A\&A, 449, 1297
\bibitem[Bizzocchi et al.(2004)]{bizzocchi04}
Bizzocchi, L., Degli Esposti, C., Botschwina, P.~2004, JMoSp, 225, 145
\bibitem[Boogert et al.(2015)]{boogert}
Boogert, A.C.A., Gerakines, P.A., Whittet, D.C.B.~2015, ARA\&A, 53, 541
\bibitem[Broten et al.(1978)]{broten}
Broten, N.W., Oka, T., Avery, L.W., MacLeod, J.M., Kroto, H.W.~1978
\bibitem[Brown \& Charnely(1990)]{bec}
Brown, P.D. \& Charnley, S.B.~1990, MNRAS, 244, 432
\bibitem[Caselli et al.(2012)]{caselli12}
Caselli, P., Keto, E., Bergin, E.A., Tafalla, M., Aikawa, Y. et al.~2012, ApJL, 759, 37
\bibitem[Ceccarelli et al.(2014)]{ceccarelli14}
Ceccarelli, C., Dominik, C., L\'opez-Sepulcre, A., Kama, M., et al.~2014, ApJL, 790, 1
\bibitem[Cernicharo \& Guelin(1996)]{ceg}
Cernicharo, J. \& Gu\'elin, M.~1996, A\&A, 309, L27
\bibitem[Chapillon et al.(2012)]{chapillon}
Chapillon, E., Dutrey, A., Guilloteau, S., Pi\'etu, V., Wakelam, V. et al.~2012, ApJ, 756, 58
\bibitem[Cheikh(2012)]{cheikh}
Cheikh, Sid Ely, PhD thesis, Universit\'e de Rennes, 2012
\bibitem[Chini et al.(1997)]{chini}
Chini, R., Reipurth, B., Ward-Thompson, D., Bally, J., Nyman, L.-A., Sievers, A., Billawala, Y.~1997, ApJ, 474, L135
\bibitem[Cleeves et al.(2014)]{cleeves}
Cleeves, L.I., Bergin, E.A., Alexander, C.M.O.'D., Du, F., Graninger, D. et al.~2014, Science, 345, 1590
\bibitem[Cordiner et al.(2011)]{cordiner11}
Cordiner, M.A., Charnley, S.B., Buckle, J.V., Walsh, C., Millar, T.J.~2011, ApJ, 730, L18
\bibitem[Cordiner et al.(2012)]{cordiner12}
Cordiner, M.A., Charnley, S.B., Wirstr\"{o}m, E.S., Smith, R.G.~2012, ApJ, 744, 131
\bibitem[Crimier et al.(2009)]{crimier}
Crimier, N., Ceccarelli, C., Lefloch, B., Faure, A.~2009, A\&A, 506, 1229 
\bibitem[Dauphas \& Chaussidon(2011)]{dec}
Dauphas, N. \& Chaussidon, M. 2011, AREPS, 39, 351
\bibitem[Fournier(2014)]{fournier14}
Fournier, M.,~PhD thesis, Universit\'e de Rennes, 2014
\bibitem[Friesen \& Pineda et al.(2017)]{friesen17}
Friesen, R.K., Pineda, J.E., Rosolowsky, E., Alves, F., Chac\'on-Tanarro, A., et al.~2017, arXiv: 170406318
\bibitem[Furlan et al.(2014)]{furlan}
Furlan, E., Megeath, S.T., Osorio, M., Stutz, A.M., Fischer, W.J. et al.~2014, ApJ, 786, 26
\bibitem[Gounelle et al.(2013)]{gounelle}
Gounelle, M, Chaussidon, M., Rollion-Bard, C.~2013, ApJ, 763, L33
\bibitem[Graninger et al.(2014)]{graninger}
Graninger, D.M., Herbst, E., \"{O}berg, K.I., Vasyunin, A.I.~2014, ApJ, 787, 74
\bibitem[Hassel et al.(2008)]{hassel}
Hassel, G.E., Herbst, E., Garrod, R.T.~2008, ApJ, 681, 1385
\bibitem[Hirota et al.(2004)]{hirota}
Hirota, T., Maezawa, H., Yamamoto, S.~2004, ApJ, 617, 399
\bibitem[Hirota et al.(2007)]{hirota07}
Hirota, T., Bushimata, T., Choi, Y.K., Honma, M., Imai, H.~2007, PASJ, 59, 897
\bibitem[Hogerheijde et al.(2012)]{hogerheijde}
Hogerheijde, M.R., Bergin, E.A., Brinch, C., Cleeves, L.I., Fogel, J.K.J. et al.~2012, Proceedings of the symposium "From Atoms to Pebbles: Herschel's view of Star and Planet Formation", Grenoble, France, March 20-23 2012, Eds.: J.-C. Augereau
\bibitem[Jaber et al.(2017)]{jaber}
Jaber, A.A., Ceccarelli, C., Kahane, C., Viti, S., Balucani, N., et al.~2017, A\&A, 597, 40
\bibitem[Jenkins(2009)]{jenkins}
Jenkins, E.B.~2009, ApJ, 700, 1299
\bibitem[J\l{o}rgensen \& Van Dishoeck(2010)]{jorgensen}
J\l{o}rgensen, J.K. \& van Dishoeck, E.F.~2010, ApJL, 710, 72
\bibitem[Kong et al.(2015)]{kong}
Kong, S., Caselli, P., Tan, J.C., Wakelam, V., Sipil\"{a}, O. et al.~2015, ApJ, 804, 98
\bibitem[Loison et al.(2014)]{loison}
Loison, J.-C., Wakelam, V., Hickson, K.M., Bergeat, A., Mereau, R.~2014, MNRAS, 437, L430
\bibitem[Lopez-Sepulcre et al.(2013a)]{lopez13}
L\'opez-Sepulcre, A., Taquet, V., S\'anchez-Monge, \'A., Ceccarelli, C., Dominik, C. et al.~2013a, A\&A, 556, 62
\bibitem[Lopez-Sepulcre et al.(2013b)]{lopez13b}
L\'opez-Sepulcre, A., Kama, M., Ceccarelli, C., Dominik, C., Caux, E., et al.~2013b, A\&A, 549, 114
\bibitem[Mauersberger et al.(1990)]{mauersberger}
Mauersberger, R., Henkel, C., Sage, L.J.~1990, A\&A, 236, 63
\bibitem[Menten et al.(2007)]{menten}
Menten, K.M., Reid, M.J., Forbrich, J., Brunthaler, A.~2007, A\&A, 474, 515
\bibitem[Mezger et al.(1990)]{mezger}
Mezger, P.G., Zylka, R., Wink, J.E.~1990, A\&A, 228, 95	
\bibitem[Milam et al.(2005)]{milam2005}
Milam, S.S.N., Savage, C., Brewster, M.A., Ziurys, L.M., Wyckoff, S.~2005, ApJ, 634, 1126
\bibitem[M\"{u}ller et al.(2001)]{muller01}
M\"{u}ller, H.S.P., Thorwirth, S., Roth, D.A., \& Winnewisser, G.~2001, A\&A, 370, L49
\bibitem[M\"{u}ller et al.(2005)]{muller05}
M\"{u}ller, H.S.P, Schl\"{u}der, F., Stutzki, J., \& Winnewisser, G.~2005, J. Mol. Struct., 742, 215
\bibitem[Mumma \& Charnley(2011)]{mec2011}
Mumma, M.J. \& Charnley S.B.~2011, ARA\&A, 49, 471
\bibitem[Noble et al.(2015)]{noble}
Noble, J.A., Theule, P., Congiu, E., Dulieu, F., Bonnin, M. et al.~2015, A\&A, 576, 91
\bibitem[\"{O}berg et al.(2015)]{oberg}
\"{O}berg, K.I., Guzm\'an, V.V., Furuya, K., Qi, C., Aikawa, Y. et al.~2015, Nature, 520, 198
\bibitem[Ossenkopf \& Henning(1994)]{oeh}
Ossenkopf, V., \& Henning, Th. 1994, A\&A, 291, 943
\bibitem[Podio et al.(2013)]{podio13}
Podio, L., Kamp, I., Codella, C., Cabrit, S., Nisini, B. et al.~2013, ApJL, 766, 5
\bibitem[Reipurth et al.(1999)]{reipurth}
Reipurth, B., Rodr\'iguez, L.F., Chini, R.~1999, AJ, 118, 983
\bibitem[Ruffle et al.(1997)]{ruffle}
Ruffle, D.P., Hartquist, T.W., Taylor, S.D., Williams, D.A.~1997, MNRAS, 291, 235
\bibitem[Sakai et al.(2008)]{sakai08}
Sakai, N., Sakai, T., Hirota, T., Yamamoto, S.~2008, ApJ, 672, 371
\bibitem[Shimajiri et al.(2008)]{shimajiri}
Shimajiri, Y., Takahashi, S., Takakuwa, S., Saito, M., Kawabe, R.~2008, ApJ, 683, 255
\bibitem[Thorwirth et al.(2000)]{thorwirth}
Thorwirth, S., M\"{u}ller, H.S.P., Winnewisser, G.~2000, JMoSp, 204, 133
\bibitem[Vuitton et al.(2007)]{vuitton}
Vuitton, V., Yelle, R.V., McEwan, M.J.~2007, Icarus, 191, 722
\bibitem[Wakelam et al.(2014)]{wakelam2014}
Wakelam, V., Vastel, C., Aikawa, Y., Coutens, A., Bottinelli, S., Caux, E.~2014, MNRAS, 445, 2854
\bibitem[Wakelam et al.(2010)]{wakelam2010}
Wakelam, V., Herbst, E., Le Bourlot, J., Hersant, F., Selsis, F., Guilloteau, S.~2010, A\&A, 517, 21 
\end{thebibliography}
\end{document}